\documentclass[12pt, fleqn, a4paper]{article} 
\usepackage{hyperref}
\hypersetup{pdfstartview = FitH,
            pdfborder    = {0 0 0.5},
            colorlinks   = false
}
 \usepackage{epsfig, subcaption}
 \usepackage{tikz}
 \usetikzlibrary{arrows.meta}
 \usepackage{color}
 \usepackage{clshan-math, clshan-dm-dd}
 \newcommand{\VchiLab}    {{\bf v}_{\chi, {\rm Lab}}}
 \newcommand{\vchiLab}    {     v _{\chi, {\rm Lab}}}
 \newcommand{\Vchiout}    {{\bf v}_{\chi_{\rm out}}}
 \newcommand{\xchi}       {{\bf X}_{\chi_{\rm in}}}
 \newcommand{\ychi}       {{\bf Y}_{\chi_{\rm in}}}
 \newcommand{\zchi}       {{\bf Z}_{\chi_{\rm in}}}
 \newcommand{\thetaNRchi} {\theta_{\rm N_R, \chi_{in}}}
 \newcommand{\chiin}      {\chi_{\rm in}}
 \newcommand{\Echi}       {     E _{\chi}}
 \newcommand  {\Ql}       {Q_{\rm l}}
 \newcommand  {\Qh}       {Q_{\rm h}}
 \newcommand  {\vminl}    {v_{\rm min, l}}
 \newcommand  {\vminh}    {v_{\rm min, h}}
 \newcommand{\rmF}        {\rmXA{F}  {19}}
 \newcommand{\rmAr}       {\rmXA{Ar} {40}}
 \newcommand{\rmGe}       {\rmXA{Ge} {73}}
 \newcommand{\rmXe}       {\rmXA{Xe}{129}}
 \newcommand{\rmW}        {\rmXA{W} {183}}
\newcommand{\InsertPlotD} [4] {
\begin{figure} [t!]
\begin{center}
 \begin{subfigure} [c] {8.25 cm}
  \includegraphics [width = 8.25 cm] {#1}
 \caption{}
 \end{subfigure}
 \hspace{0.1 cm}
 \begin{subfigure} [c] {8.25 cm}
  \includegraphics [width = 8.25 cm] {#2}
 \caption{}
 \end{subfigure}
\end{center}
\caption{
 #4
}
\label{fig:#3}
\end{figure}
}
\newcommand{\OnlinePlotNRchieta} [3] {
\href{http://www.tir.tw/phys/hep/dm/amidas-2d/amidas-2d.php%
      ?amidas_2D_function=NR_theta%
      &mode_NR=#1_theta%
      &frame=chi%
      &mode_animation=target%
      &mchi=#2%
      &period=periodA%
      &event_No=500}
     {#3}%
}
\newcommand{\OnlinePlotNRangEq} [3] {
\href{http://www.tir.tw/phys/hep/dm/amidas-2d/amidas-2d.php%
      ?amidas_2D_function=NR_ang%
      &mode_NR=#1_ang%
      &frame=Eq%
      &mode_animation=mchi%
      &target=#2%
      &period=periodA%
      &event_No=500}
     {#3}%
}
\begin{document}
\thispagestyle{empty}
\begin{flushright}
 October 2021
\end{flushright}
\begin{center}
{\large\bf
 Some Thoughts on (the Incompleteness of)
 the (Double) \\ Differential Event Rates
 for Elastic WIMP--Nucleus Scattering                   \\ \vspace{0.1 cm}
 in (Directional) Direct Dark Matter Detection Physics} \\
\vspace*{0.7 cm}
 {\sc Chung-Lin Shan}                                   \\
\vspace{0.5 cm}
 {\small\it
  Preparatory Office of
  the Supporting Center for
  Taiwan Independent Researchers                        \\ \vspace{0.05 cm}
  P.O.BOX 21 National Yang Ming Chiao Tung University,
  Hsinchu City 30099, Taiwan, R.O.C.}                   \\~\\~\\
 {\it E-mail:} {\tt clshan@tir.tw}
\end{center}
\vspace{2 cm}
\begin{abstract}

 In this paper,
 we revisit the expressions for
 the (double) differential event rates
 for elastic WIMP--nucleus scattering
 and
 discuss some unusual thoughts on
 (the incompleteness of)
 the uses of these expressions
 in (directional) direct Dark Matter detection physics.
 Several not--frequently mentioned (but important) issues
 will be argued and demonstrated in detail.

\end{abstract}
\clearpage
\section{Introduction}

 Direct Dark Matter (DM) detection experiments
 aiming to observe scattering signals of
 Weakly Interacting Massive Particles (WIMPs)
 off target nuclei
 by measuring recoil energies
 deposited in an underground detector
 would still be the most reliable experimental strategy
 for identifying Galactic DM particles
 and determining their properties
  \cite{SUSYDM96, Mayet16, Schumann19, Baudis20, Vahsen21, Cooley21}.
 For either
 providing exclusion limits
 with null results
 or reconstructing WIMP properties
 with positive scattering events,
 an estimate of
 the total or the differential event rate
 for elastic WIMP--nucleus scattering
 is essential and crucial
 \cite{SUSYDM96, Schumann19, Baudis20, Cooley21}.

 During our works on
 developing
 the double Monte Carlo scattering--by--scattering simulation procedure of
 3-dimensional elastic WIMP--nucleus scattering events
 \cite{DMDDD-3D-WIMP-N}
 as well as
 studying the 3-D WIMP effective velocity distribution
 in the Galactic and the Equatorial coordinate systems
 \cite{DMDDD-fv_eff},
 we revisited the expressions for
 the (double) differential event rates
 for elastic WIMP--nucleus scattering
 used conventionally
 in (directional) direct DM detection physics.
 Several incompatibilities
 between our microscopic and the conventional macroscopic approaches
 have however been observed.

 Hence,
 in this paper,
 we would like to discuss
 these (unusual) thoughts on
 (the incompleteness of)
 the expressions for
 the (double) differential event rates.
 In Sec.~2,
 we consider at first
 the general expression for
 the differential event rate
 needed basically
 in all direct DM detection physics.
 Then
 issues of
 the expression for
 the double differential event rate
 used particularly
 in directional direct detection physics
 will be argued in Sec.~3.
 We conclude
 in Sec.~4.

%
 %
%
\section{Differential event rate}
\label{sec:dRdQ}

 In this section,
 we discuss at first
 some possible incompleteness of
 the differential event rate
 for elastic WIMP--nucleus scattering.

 In literature,
 the general expression for
 the differential event rate
 for elastic WIMP--nucleus scattering
 used in direct DM detection physics
 has been given by
 \cite{SUSYDM96, Schumann19, Baudis20, Cooley21}
\beq
     \dRdQ
  =  \frac{\rho_0}{2 \mchi \mrN^2}
     \bbigg{\sigmaSI \FSIQ + \sigmaSD \FSDQ}
     \intvminQ \bfrac{f_1(\vchiLab)}{\vchiLab} d\vchiLab
\~.
\label{eqn:dRdQ_SISD}
\eeq
 Here $R$ is the detection event rate,
 $Q$ is the nuclear recoil energy deposited in the detector,
 $\rho_0$ is the WIMP density near the Earth,
 $f_1(\vchiLab)$ is
 the one--dimensional velocity distribution function
 of the WIMPs {\em impinging} on the detector,
 $\vchiLab$ is
 the magnitude of the 3-D WIMP incident velocity
 in the laboratory frame.
 $\sigma_0^{\rm (SI, SD)}$ are
 the spin--independent/dependent (SI/SD) total cross sections
 ignoring the nuclear form factor suppressions
 and
 $F_{\rm (SI, SD)}(Q)$ indicate the elastic nuclear form factors
 corresponding to the SI/SD WIMP interactions,
 respectively.
\(
         \mrN
 \equiv  \mchi \mN / \abrac{\mchi + \mN}
\)
 is the reduced mass of
 the WIMP mass $\mchi$
 and that of the target nucleus $\mN$.
 Finally,
 $\vmin(Q)$ is
 the minimal--required incoming velocity of incident WIMPs
 that can deposit the energy $Q$ in the detector:
\beq
     \vmin(Q)
  =  \sfrac{\mN}{2 \mrN^2} \~ \sqrt{Q}
\~.
\label{eqn:vmin}
\eeq
\subsection{Recoil--angle dependence of the recoil energy}
\label{sec:Q_eta}
\begin{figure} [t!]
\begin{center}
\begin{tikzpicture}
      [vector/.style = {-Stealth,
                        line width = 2 pt},
       dashed/.style = {dash pattern = on 0.225 cm off 0.125 cm,
                        line width = 1.5 pt},
       angle/.style  = {Stealth-Stealth, line width = 0.75 pt}]
 \draw [very thick, color = white]
       (0, 0) rectangle (12 ,  6.8 );
%
%
 \draw [fill]      (  1.5  , 3.3  ) circle [radius = 0.1 ]
                  +(  0    ,-0.1  ) [below] node {\large $\mchi$};
 \draw [vector]    (  1.75 , 3.3  ) -- (  6.75 , 3.3  );
 \node [below]  at (  4.25 , 3.2  ) {\large $\VchiLab$};
%
%
 \draw [fill]      (  7    , 3.3  ) circle [radius = 0.15]
                  +(- 0.15 , 0.1  ) [above] node {\large $\mN$};
%
%
 \draw [vector]    (  7    , 3.3  ) +( 0.263, 0.425) -- +( 2.167, 3.5  );
 \node [below]  at (  7.6  , 5.8  ) {\large $\Vchiout$};
 \draw [angle]     (  7.75 , 3.3  ) arc [radius = 0.75 , start angle =   0, end angle =  58.237];
 \node          at (  7.9  , 3.85 ) {\large $\zeta$};
%
%
 \draw [vector]    (  7    , 3.3  )                  -- +( 1.25 ,-1.5  );
 \node [below]  at (  8.7  , 1.85 ) {\large ${\bf v}_{\rm NR}$};
 \draw [angle]     (  7.75 , 3.3  ) arc [radius = 0.75 , start angle =   0, end angle = -50.194];
 \node          at (  7.9  , 2.75 ) {\large $\eta$};
%
%
 \draw [dashed]    (  7.25 , 3.3  ) -- +( 2.325, 0    );
 \node [right]  at (  9.6  , 3.2  ) {\large $\zchi$};
%
 \draw [dashed]    (  7    , 3.05 ) -- +( 0    ,-2.325);
 \node [below]  at (  7.5  , 0.7  ) {\normalsize $\xchi$--$\ychi$ plane};
 \draw [angle]     (  7    , 2.25 ) arc [radius = 1.05, start angle = -90, end angle = -50.194];
 \node          at (  7.65 , 1.85 ) {\normalsize $\thetaNRchi$};
\end{tikzpicture}
\vspace{-0.25 cm}
\end{center}
\caption{
 The sketch of a WIMP--nucleus ($\chi$--N) scattering event
 with a WIMP incident velocity of $\vchiLab$.
 $\zeta$ is
 the scattering angle of
 the outgoing WIMP,
 $\eta$ is
 the recoil angle of
 the scattered target nucleus,
 and
 $\thetaNRchi = \pi / 2 - \eta$ is
 the elevation of
 the recoil direction of the scattered nucleus
 (the equivalent recoil angle)
 in the incoming--WIMP ($\chiin$) coordinate system
 \cite{DMDDD-3D-WIMP-N}.
 The $\zchi$--axis is defined as
 the direction of the incoming velocity of
 the incident WIMP
 $\VchiLab$.
}
\label{fig:chi-N}
\end{figure}
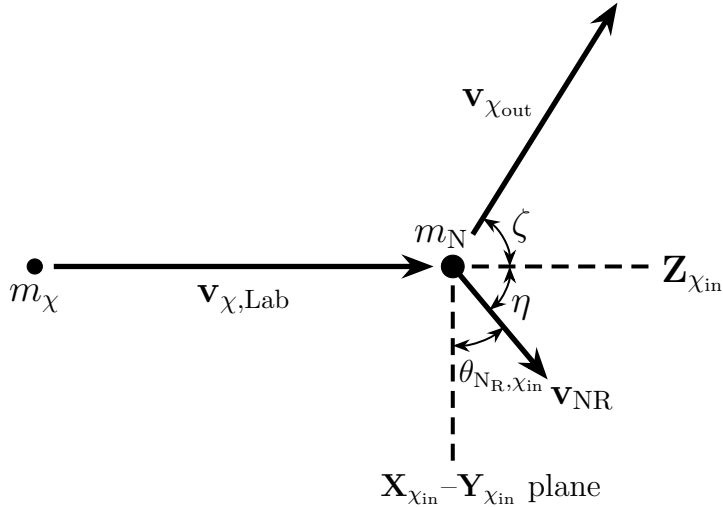

 In Fig.~\ref{fig:chi-N}
 we sketch
 a WIMP--nucleus ($\chi$--N) scattering event
 with a WIMP incident velocity of $\vchiLab$.
 $\zeta$ is
 the scattering angle of
 the outgoing WIMP,
 $\eta$ is
 the recoil angle of
 the scattered target nucleus,
 and
 $\thetaNRchi = \pi / 2 - \eta$ is
 the elevation of
 the recoil direction of the scattered nucleus
 (the ``equivalent'' recoil angle)
 in the incoming--WIMP ($\chiin$) coordinate system
 \cite{DMDDD-3D-WIMP-N}.
 While
 $\mchi$ indicates
 (the mass of) the incident WIMP,
 $\mN$ is
 (the mass of) the scattered target nucleus
 in our detector.

 The kinetic energy of the incident WIMP
 with the incoming velocity of $\VchiLab$
 in the laboratory (detector rest) coordinate system
 can be given by
\beq
     \Echi
  =  \frac{1}{2} \mchi |\VchiLab|^2
  =  \frac{1}{2} \mchi  \vchiLab ^2
\~.
\label{eqn:Echi}
\eeq
 Then
 the recoil energy of the scattered target nucleus
 can be expressed as
 a function of the recoil angle $\eta$ as
 \cite{Billard09,
       OHare14,
       Mayet16}
\beq
     Q(\vchiLab, \eta)
  =  \bbrac{\frac{4 \mchi \mN}{(\mchi + \mN)^2} \~ \cos^2(\eta)}
     \Echi
  =  \afrac{2 \mrN^2}{\mN} \vchiLab^2
     \cos^2(\eta)
\~,
 \label{eqn:QQ_eta}
\eeq
 where
 $\mrN$ is the reduced mass
 given above.

\begin{figure} [p!]
\begin{center}
 \begin{subfigure} [c] {13 cm}
  \includegraphics [width = 13 cm] {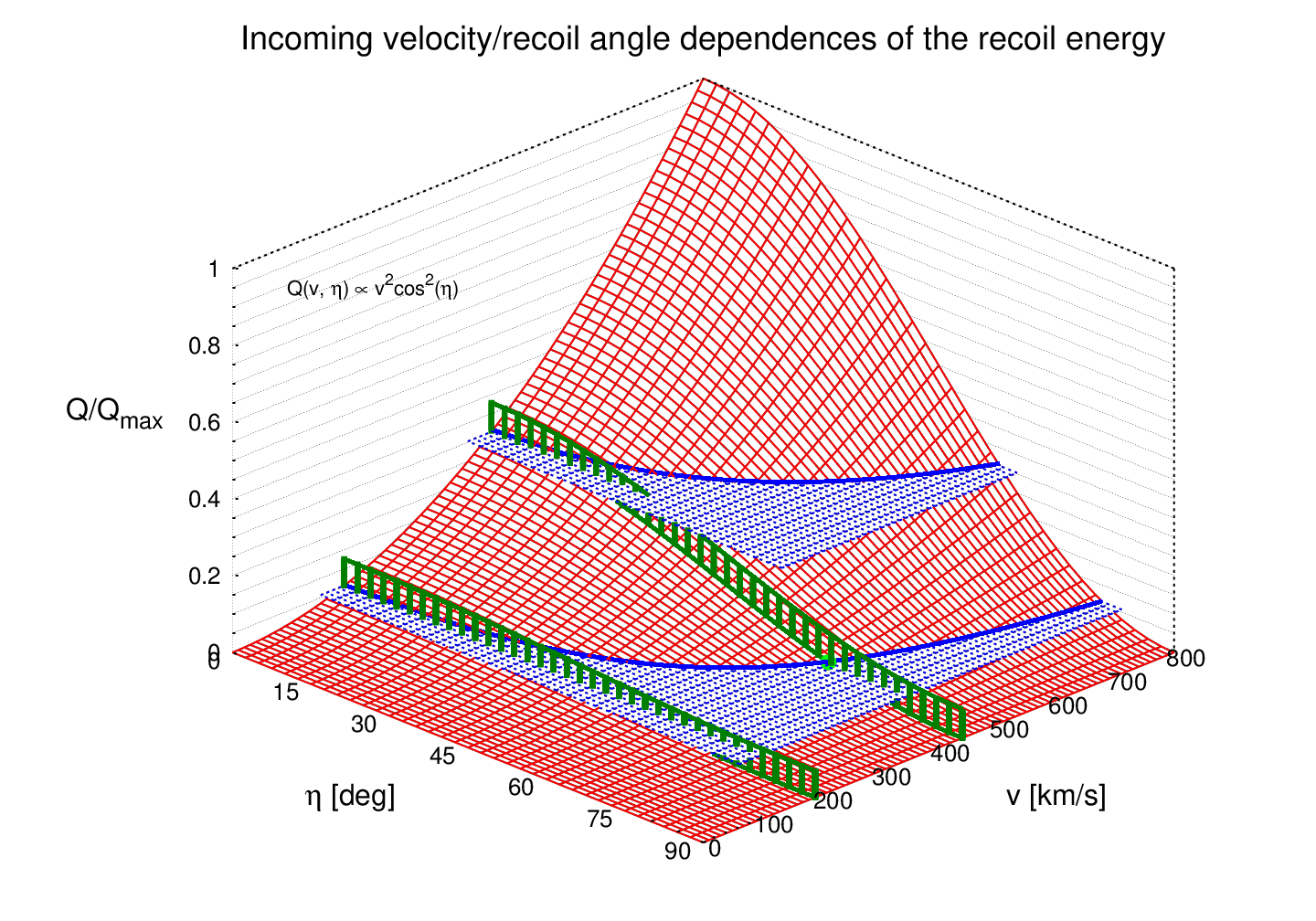}
 \caption{}
 \end{subfigure}
 \\
 \vspace{0.75 cm}
 \begin{subfigure} [c] {13 cm}
  \includegraphics [width = 13 cm] {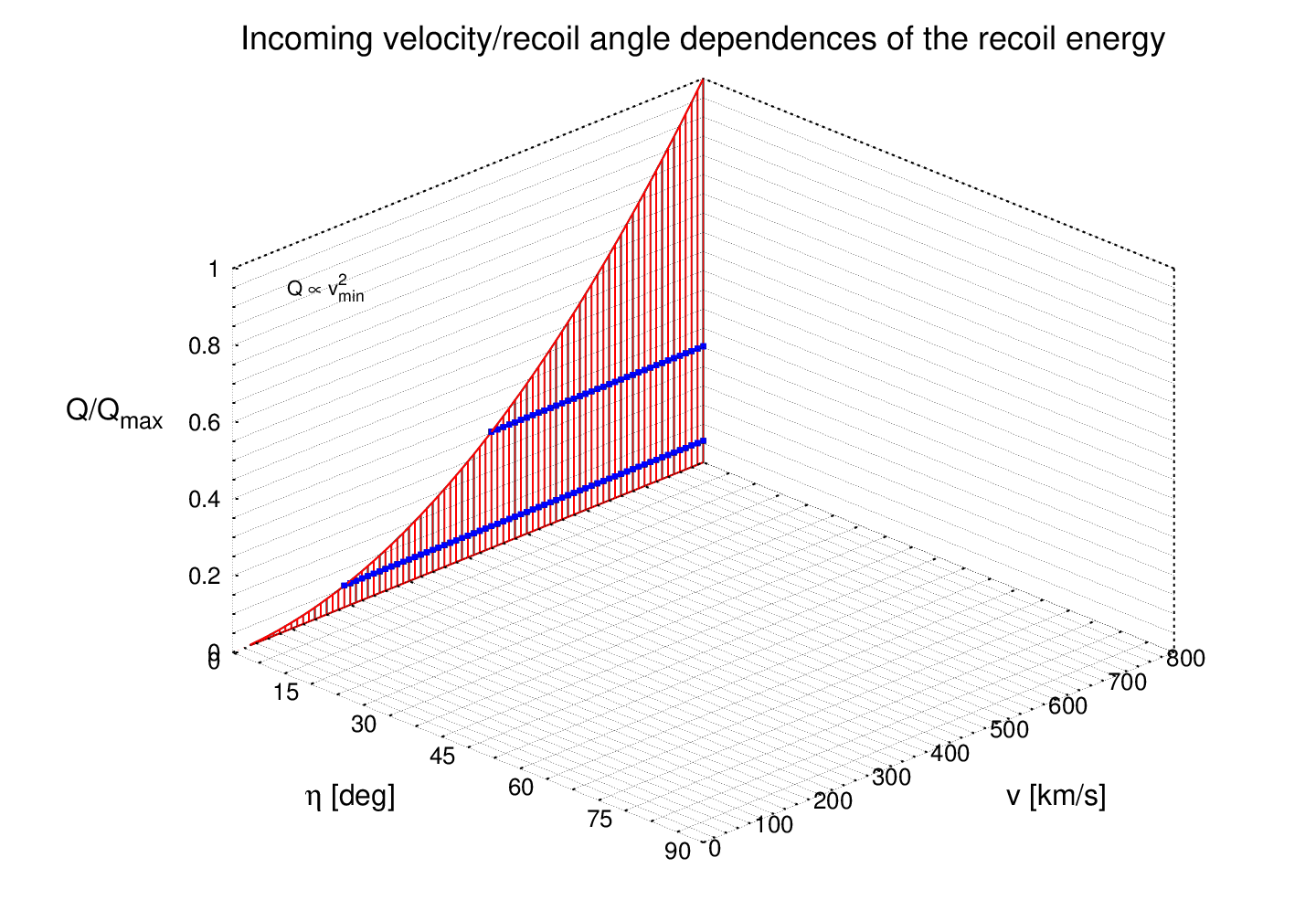}
 \caption{}
 \end{subfigure}
\\
\vspace{-20.25 cm}
\begin{picture}(13   , 20.25)
\color{blue}
\put(10.9 , 14.42){\makebox(0.6 , 0.4 ){$\Ql$}}
\put( 9.9 , 15.82){\makebox(0.6 , 0.4 ){$\Qh$}}
\put( 2.35, 14.65){\makebox(1   , 0.3 ){$\vminl$}}
\put( 3.8 , 16.2 ){\makebox(1   , 0.3 ){$\vminh$}}
\put( 7.05,  5.3 ){\makebox(0.6 , 0.4 ){$\Ql$}}
\put( 7.05,  6.25){\makebox(0.6 , 0.4 ){$\Qh$}}
\put( 2.35,  3.95){\makebox(1   , 0.3 ){$\vminl$}}
\put( 3.8 ,  5.5 ){\makebox(1   , 0.3 ){$\vminh$}}
\end{picture}
\end{center}
\caption{
 (a)
 The 2-D dependence of the recoil energy on
 the WIMP incident velocity $\vchiLab$
 and the recoil angle $\eta$
 given in Eq.~(\ref{eqn:QQ_eta}).
 (b)
 The projection of the recoil--energy surface
 on the $\eta = 0$ plane.
 See the text for detailed descriptions and discussions.
}
\label{fig:Qeta-v_l-v_h-Q_l-Q_h}
\end{figure}
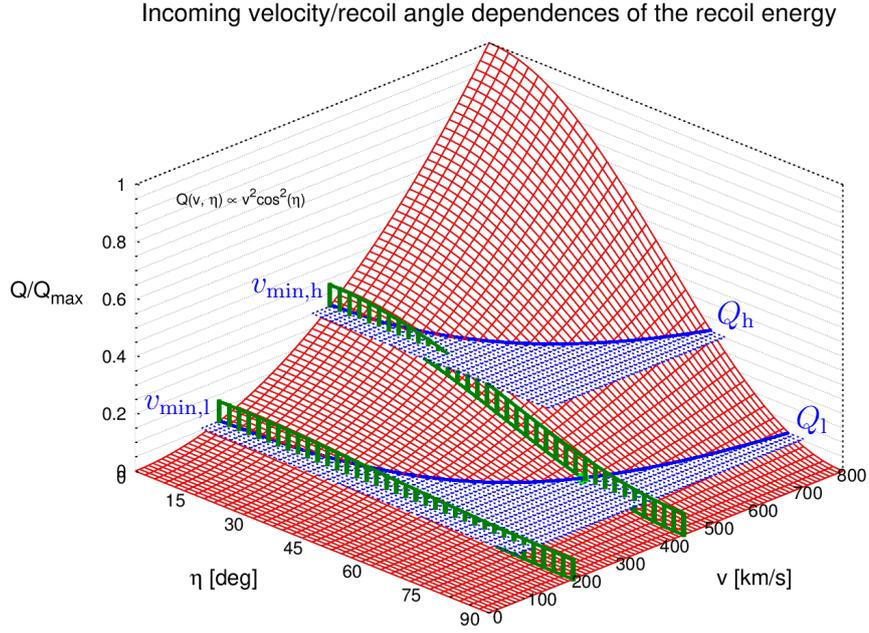
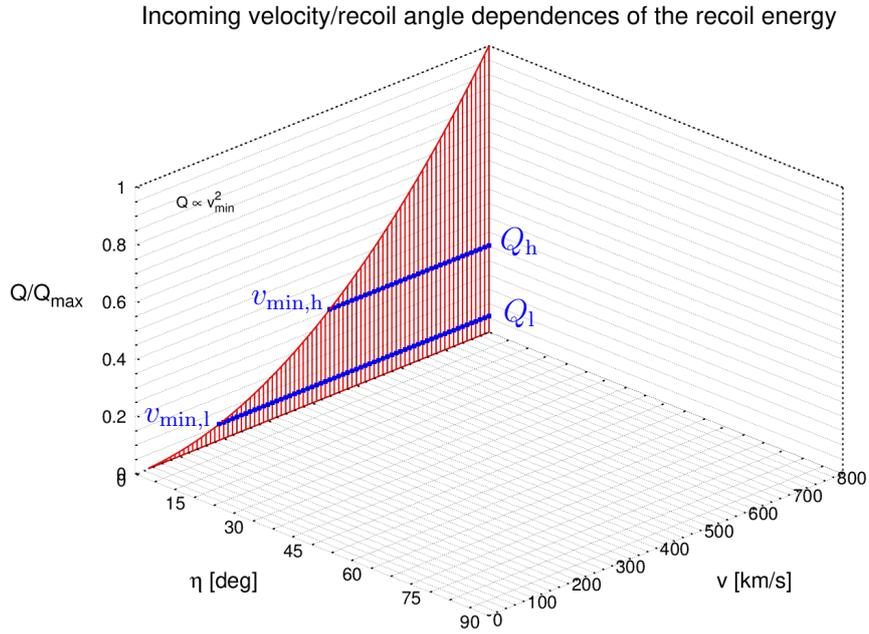

 Conventionally,
 for a given recoil energy $\Qh$
 (see Fig.~\ref{fig:Qeta-v_l-v_h-Q_l-Q_h}(b)),
 people use Eq.~(\ref{eqn:QQ_eta})
 with the minimal recoil angle of $\eta = 0$
 (namely,
  Eq.~(\ref{eqn:vmin}))
 to define the minimal--required incoming velocity
 $\vminh \equiv \vmin(\Qh)$
 appearing as the lower bound of the integral
 in Eq.~(\ref{eqn:dRdQ_SISD}).
 Then,
 for estimating the differential event rate $\abrac{dR / dQ}_{Q = \Qh}$,
 we integral over
 the 1-D velocity distribution function $f_1(\vchiLab)$
 (divided by $\vchiLab$)
 from $\vminh$
 to a maximal cut--off velocity of incident WIMPs
 in the Equatorial/laboratory coordinate systems
 ($v_{\chi, {\rm cutoff}} \simeq 800$ km/s)
 by assuming implicitly that
 {\em the velocity distribution of
  the ``scattering'' WIMPs
  (moving with velocities larger than $\vminh$)
  is exactly identical to
  the 1-D velocity distribution of
  ``incident'' halo WIMPs},
  $f_1(\vchiLab)$,%
\footnote{
 See Refs.~\cite{DMDDD-fv_eff, DMDDD-v_theta}
 for detailed discussions and demonstrations against
 this assumption.
}
 and ignoring
 the dependence of the recoil energy
 on the recoil angle $\eta$
 through Eq.~(\ref{eqn:QQ_eta}),
 as along the upper solid (blue) line
 shown in Fig.~\ref{fig:Qeta-v_l-v_h-Q_l-Q_h}(b).
 This implies in practice that,
 for another recoil energy $\Ql < \Qh$
 with the corresponding minimal WIMP--incoming velocity
 $\vminl \equiv \vmin(\Ql)$,
 the predicted contribution of incident WIMPs
 moving with velocities
 $\vchiLab \ge \vminh$
 to the differential event rate
 $\abrac{dR / dQ}_{Q = \Ql}$
 would exactly be equal to
 that to $\abrac{dR / dQ}_{Q = \Qh}$.
 Additionally,
 for different target nuclei,
 as long as the lower bounds of the integral are equal
 (to e.g.~$\vminh$ or $\vminl$),
 the predicted contributions of the integral
 to the differential event rates
 (around different recoil energies!)
 would also be equal.

 However,
 since the recoil energy depends on
 not only the WIMP incident velocity $\vchiLab$
 but also the nuclear recoil angle $\eta$,
 we would need to consider a 2-D surface
 as shown in Fig.~\ref{fig:Qeta-v_l-v_h-Q_l-Q_h}(a).
 For the given recoil energy $\Qh$,
 the integral in the differential event rate
 $\abrac{dR / dQ}_{Q = \Qh}$
 should go along the upper (dark--blue) intersection path of
 the (red) $Q(\vchiLab, \eta)$ surface
 and the (blue) ``equal--recoil--energy--$\Qh$'' plane.
 This means that
 the contribution of the integral
 in the range of $\vchiLab \ge \vminh$ to
 $\abrac{dR / dQ}_{Q = \Qh}$
 would be totally different from
 the contribution of the integral to
 $\abrac{dR / dQ}_{Q = \Ql}$
 (the lower (dark--blue) intersection path).
 This would then make (the expression for)
 the estimate of
 the differential event rate
 (much) more complicated.
 Below,
 in the rest part of this section,
 we provide from the theoretical point of view
 the most important (but may not be usually considered) arguments for it%
\footnote{
 A detailed investigation on
 the (target and WIMP--mass dependent)
 incident velocity--recoil angle distribution for
 elastic WIMP--nucleus scattering
 in different (small) energy windows
 will be announced later
 \cite{DMDDD-v_theta}.
}.

 First of all,
 as shown in Fig.~\ref{fig:Qeta-v_l-v_h-Q_l-Q_h}(a),
 for the given WIMP incident velocity
 $\vminh$
 (the upper (dark--green) vertical fence),
 one could consider reversely
 the corresponding recoil energy $\Qh$
 as the ``maximal transferable'' recoil energy
 to the scattered target nucleus:
\beq
         Q_{\rm max}(\vminh)
 \equiv  \afrac{2 \mrN^2}{\mN} \vminh^2
  =      \Qh
\~.
 \label{eqn:Qmax_vchiLab}
\eeq
 Then,
 for any recoil angle $\eta > 0$,
 the larger the recoil angle $\eta$
 (along downwards the upper (dark--green) vertical fence),
 the smaller the corresponding recoil energy
 $Q(\vchiLab = \vminh, \eta > 0)$.
 As will be discussed in detail in Sec.~\ref{sec:FQ},
 this implies
 a {\em weaker} cross section (nuclear form factor) suppression
 and thus possibly a {\em larger} scattering probability.
 Hence,
 incident WIMPs moving with $\vchiLab = \vminh$
 would (much) more likely scatter off a target nucleus
 with a larger recoil angle
\(
     \eta^{\ast}
  =  \cos^{-1}\abrac{\sqrt{\Ql / \Qh}}
\)
 (the green point)
 and thus transfer
 the given recoil energy $\Ql$.
 This indicates
 from so far only the nuclear--form--factor point of view
 clearly that
 head--on scattering events
 should be (much) rarer than people thought before.

 In fact,
 for all $\vchiLab \ge \vminh$,
 it would always be more possible
 to observe scattering events
 with lower recoil energy $\Ql$ and larger recoil angles
 than those with higher energy $\Qh$
 and smaller recoil angles.
 Hence,
 it would be pretty hard to believe that
 the contributions of the integrals
 along two intersection paths of
 the equal--recoil--energy--$\Qh$/$\Ql$ planes
 from $\vchiLab = \vminh$ to
 the cut--off WIMP incident velocity $v_{\chi, {\rm cutoff}}$
 to the differential event rates
 $\abrac{dR / dQ}_{Q = \Qh}$ and $\abrac{dR / dQ}_{Q = \Ql}$
 would be identical%
\footnote{
 In Ref.~\cite{DMDDD-v_theta},
 we will demonstrate that
 the ``effective'' velocity distributions of
 scattering WIMPs
 inducing recoil energies
 in different (small) energy windows
 would be clearly different
 and (much) more complicated than
 target and WIMP--mass dependent.
}.

 Moreover,
 as will be argued in more details
 in Sec.~\ref{sec:dsigma_eta},
 the scattering probability
 at different incident velocity--recoil angle ($\vchiLab$--$\eta$) combinations
 depends partially on
 the nuclear form factor,
 which depends on detector materials
 and the WIMP mass.
 Thus,
 for estimating the differential event rate $dR / dQ$,
 the form factors $F_{\rm (SI, SD)}(Q)$
 would need to be {\em included inside} the integral
 and handled as functions of
 the incident velocity $\vchiLab$
 and the recoil angle $\eta$
 through Eq.~(\ref{eqn:QQ_eta}).
\subsection{Cross section (nuclear form factor) suppression}
\label{sec:FQ}

 In the previous Sec.~\ref{sec:Q_eta},
 we have mentioned
 the dependence of
 the WIMP scattering probability
 on the cross section (nuclear form factor).
 In this subsection,
 we discuss
 the effects of the nuclear form factor suppression
 on the scattering probability
 in more details.

 \InsertPlotD
  {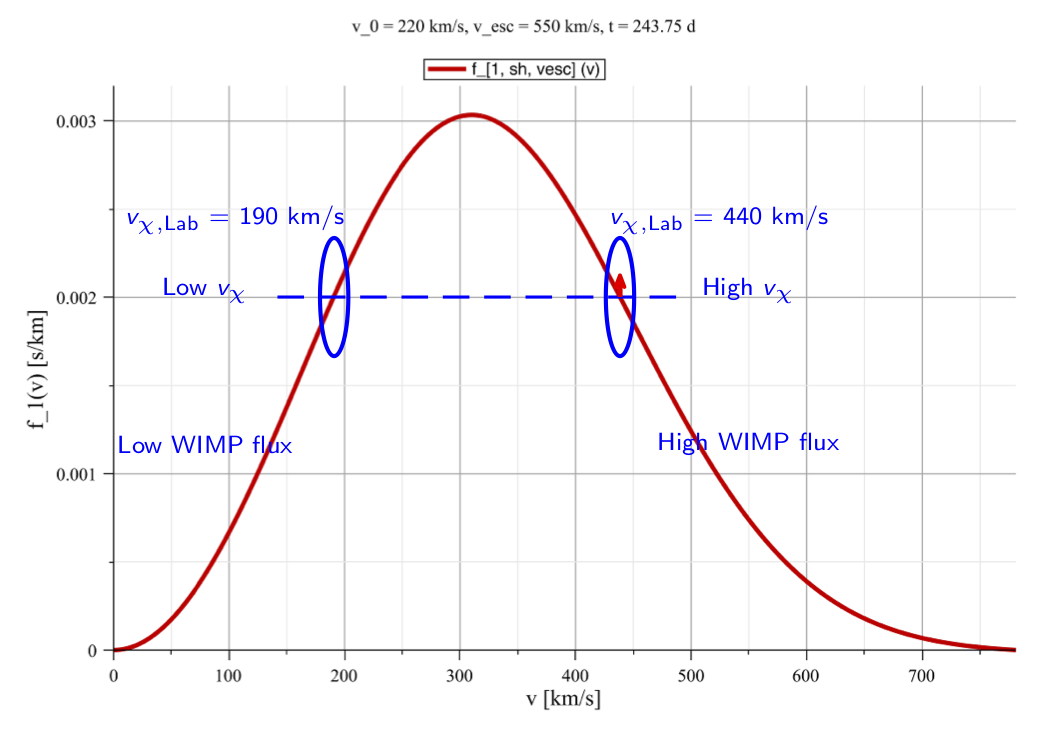}
  {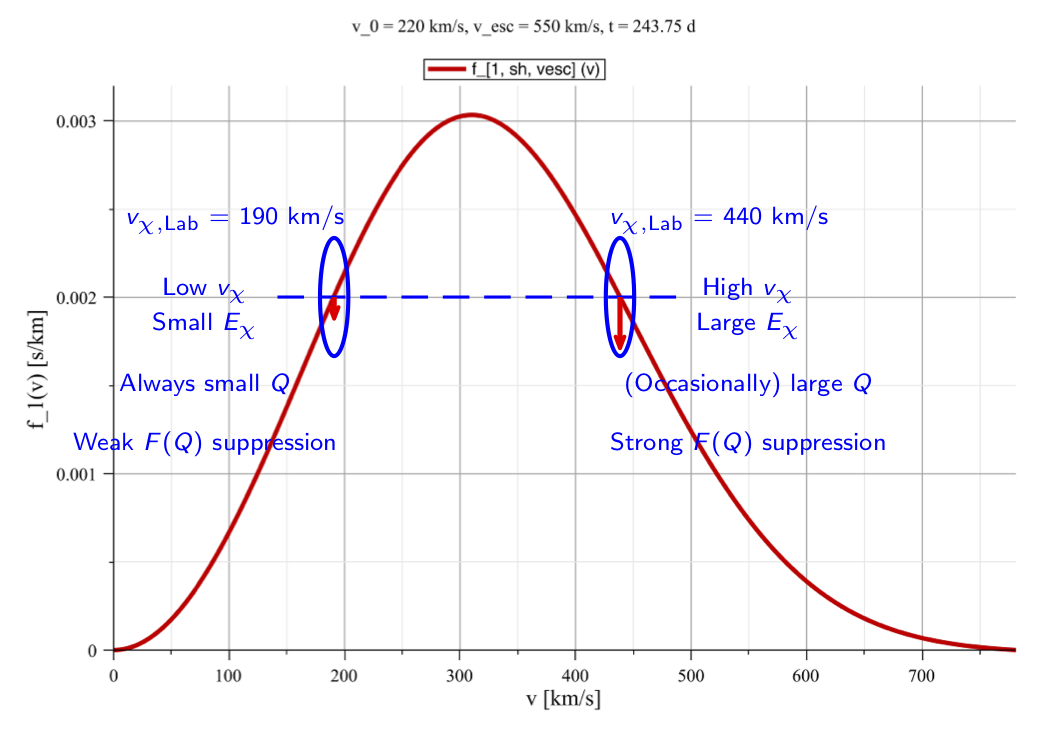}
  {fv_eff-radial}
  {Two factors
   which affect
   the dependence of the WIMP--nucleus scattering probability
   on the WIMP incident velocity:
   (a)
   the proportionality of the WIMP flux to the incident velocity,
   (b)
   the nuclear form factor suppression
   depending on the induced recoil energy,
   which is proportional to the kinetic energy of the scattering WIMP.
   The solid red curves
   are the shifted Maxwellian velocity distribution
   given in Eq.~(7) of Ref.~\cite{DMDDD-fv_eff}.
   See the text for detailed arguments.
   (Figures from Ref.~\cite{DMDDD-fv_eff}).%
   }

 Given the mass of incident WIMPs and the detector target nucleus,
 as sketched in Figs.~\ref{fig:fv_eff-radial},
 the scattering probability of
 incident halo WIMPs
 moving with different incoming velocities
 depends on two factors:
 (a)
 the WIMP flux is proportional to the incident velocity $\vchiLab$,
 and
 (b)
 the recoil angle and thus the recoil energy
 is constrained by the nuclear form factor suppression.

 The first factor of
 the flux proportionality to the WIMP incident velocity
 (sketched in Fig.~\ref{fig:fv_eff-radial}(a))
 is easy to understand
 (and was already considered
  in the differential event rate (\ref{eqn:dRdQ_SISD})):
 the higher (lower) the incoming velocity of incident WIMPs,
 the more (fewer) the target nuclei
 passed by an incident WIMP
 (in a unit time)
 and thus
 the larger (smaller)
 the scattering opportunity.

 On the other hand,
 as sketched on the left--hand side
 of Fig.~\ref{fig:fv_eff-radial}(b),
 for WIMPs moving with low incident velocities
 and thus only carrying small kinetic energies,
 the maximal transferable recoil energies to target nuclei
 are small
 and the nuclear form factor suppression
 as well as
 the reduction of the scattering probability
 are weak or even negligible,
 especially
 when light nuclei are used as detector materials
 {\em or} the WIMP mass is light
 (see also Figs.~\ref{fig:FQ}).
 In contrast,
 for WIMPs moving with high incident velocities
 and thus carrying large kinetic energies
 (the right--hand side
  of Fig.~\ref{fig:fv_eff-radial}(b)),
 the transferable recoil energies to target nuclei
 are also large
 and the nuclear form factor suppression
 as well as
 the reduction of the scattering probability
 become (pretty) strong,
 especially
 when heavy nuclei are used
 {\em and} the WIMP mass is heavy.

 \InsertPlotD
  {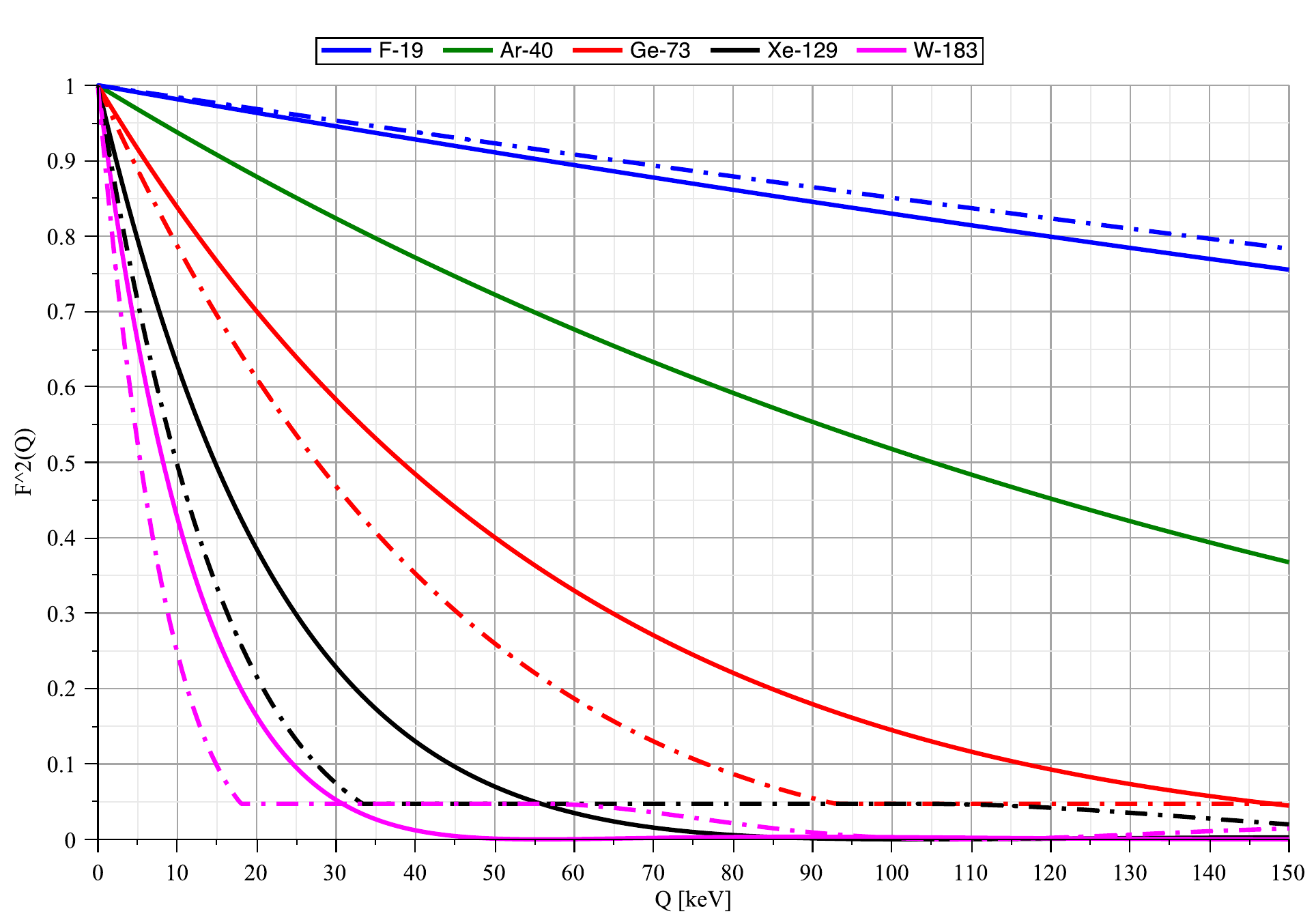}
  {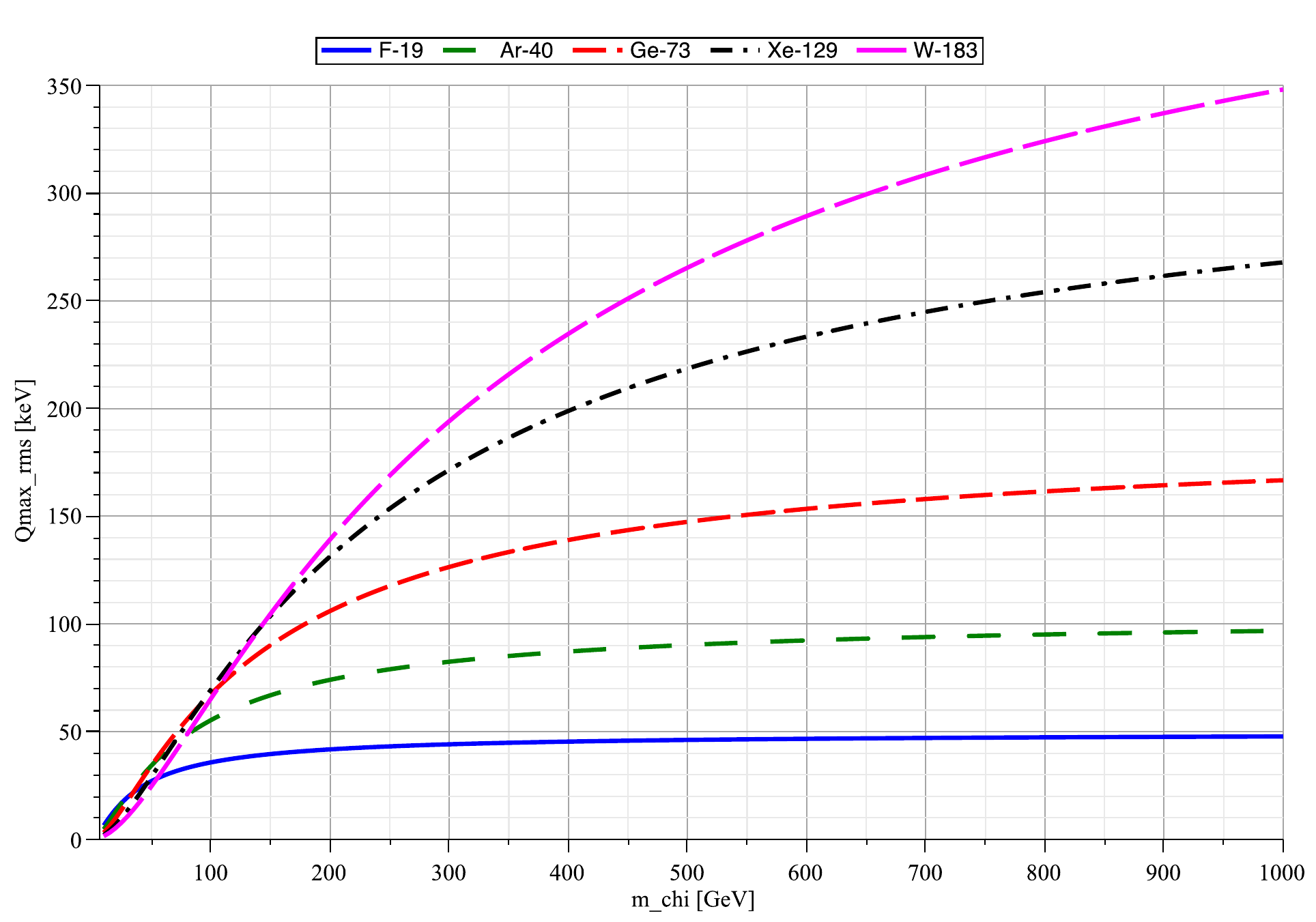}
  {FQ}
  {(a)
   Nuclear form factors of
   the $\rmF$     (blue),
   the $\rmAr$    (green),
   the $\rmGe$    (red),
   the $\rmXe$    (black),
   and the $\rmW$ (magenta) nuclei
   (adopted in our simulation package \cite{DMDDD-3D-WIMP-N})
   as functions of the recoil energy.
   The solid and the dash--dotted curves
   indicate the form factors corresponding to
   the SI and the SD cross sections,
   respectively.
   (b)
   The WIMP--mass dependence of
   the maximal recoil energy
   $Q_{\rm max, rms}$
   given by Eq.~(\ref{eqn:Qmax_rms}).
   Five frequently used target nuclei:
   $\rmF$     (solid        blue),
   $\rmAr$    (rare--dashed green),
   $\rmGe$    (dashed       red),
   $\rmXe$    (dash--dotted black),
   and $\rmW$ (long--dashed magenta)
   have been considered.
   (Figures from Ref.~\cite{DMDDD-3D-WIMP-N}).%
   }

 As readers' reference,
 in Fig.~\ref{fig:FQ}(a),
 we provide
 the recoil--energy dependence of
 the nuclear form factors corresponding to
 the SI (solid) and the SD (dash--dotted) cross sections
 adopted in our simulation package
 \cite{DMDDD-3D-WIMP-N}.
 Five frequently used target nuclei
 have been considered:
 $\rmF$     (blue),
 $\rmAr$    (green),
 $\rmGe$    (red),
 $\rmXe$    (black),
 and $\rmW$ (magenta).
 The sharply enlarged form factor suppressions
 with the increased mass of the target nucleus
 can be seen clearly.
 Moreover,
 as auxiliary material,
 Fig.~\ref{fig:FQ}(b)
 shows
 the WIMP--mass dependence of
 the maximal transferable recoil energy
 with the root--mean--square velocity
 of incident halo WIMPs
 estimated by Eq.~(40) of Ref.~\cite{DMDDD-3D-WIMP-N}:
\beq
     Q_{\rm max, rms}
  =  \afrac{2 \mrN^2}{\mN} v_{\rm rms, Lab}^2
\~,
\label{eqn:Qmax_rms}
\eeq
 for five frequently used target nuclei:
 $\rmF$     (solid        blue),
 $\rmAr$    (rare--dashed green),
 $\rmGe$    (dashed       red),
 $\rmXe$    (dash--dotted black),
 and $\rmW$ (long--dashed magenta).
 This indicates clearly that,
 due to the nuclear form factor suppression,
 especially for heavy target nuclei
 like $\rmXe$ and $\rmW$,
 even though
 the WIMPs are not very heavy
 and the incoming velocities are not very high,
 the transferable recoil energy
 would be pretty large
 and thus the scattering probability
 should be pretty strongly reduced.

 As will be discussed in detail
 in Sec.~\ref{sec:dsigma_eta}
 (and Ref.~\cite{DMDDD-v_theta}),
 this implies further that
 the deviation of
 the most frequent recoil angle
 from the zero recoil angle (head--on scattering)
 could be pretty large,
 especially
 when the target nuclei {\em and} incident WIMPs
 are heavy
 \cite{DMDDD-v_theta}.

\subsection{Recoil--angle dependence of
            the differential scattering cross section}
\label{sec:dsigma_eta}

 In literature,
 given a WIMP incident velocity $\vchiLab$,
 the differential scattering cross section $d\sigma$
 is given by
 the absolute value of the momentum transfer
 from the incident WIMP to the scattered nucleus as
 \cite{SUSYDM96, Schumann19, Baudis20, Cooley21}
\beqn
     d\sigma
 \=  \frac{1}{\vchiLab^2}
     \afrac{1}{4 \mrN^2}
     \bbigg{\sigmaSI F_{\rm SI}^2(q) + \sigmaSD F_{\rm SD}^2(q)} dq^2
     \non\\
 \=  \frac{1}{\vchiLab^2}
     \afrac{\mN}{2 \mrN^2} \bbigg{\sigmaSI \FSIQ + \sigmaSD \FSDQ} dQ
\~,
\label{eqn:dsigma_Q}
\eeqn
 where we have
\beq
     q
  =  \sqrt{2 \mN Q}
\~.
\label{eqn:qq}
\eeq
 Since,
 as shown in Eq.~(\ref{eqn:QQ_eta}),
 the recoil energy $Q$ is
 a ``one--to--one'' function of the recoil angle $\eta$
 (given the WIMP incident velocity $\vchiLab$),
 the differential cross section $d\sigma$ of
 incident halo WIMPs
 with an incoming velocity $\vchiLab$
 off target nuclei
 going into a recoil angle of $\eta \pm d\eta / 2$
 with a recoil energy of $Q \pm dQ / 2$
 can further be expressed as
 \cite{DMDDD-3D-WIMP-N}
\beq
     d\sigma
  =- \bbigg{\sigmaSI \FSIQ + \sigmaSD \FSDQ}
     \sin(2 \eta) \~ d\eta
\~.
\label{eqn:dsigma_eta}
\eeq
 Here the ``minus ($-$)'' sign indicates that
 the recoil energy $Q$ decreases
 while the recoil angle $\eta$ increases.
 Note that,
 although,
 with the zero recoil angle $\eta = 0$,
 the corresponding recoil energy $Q(\vchiLab, \eta = 0) \propto \vchiLab^2$
 is maximal,
 the differential recoil energy $dQ \propto \sin(2 \eta) = 0$
 as well as
 the differential cross section $d\sigma = 0$.
 This indicates that
 an ``exact head--on'' elastic WIMP--nucleus scattering event
 is actually {\em impossible}!

 Moreover,
 by combining Eqs.~(\ref{eqn:dsigma_eta}) and (\ref{eqn:QQ_eta}),
 one can have that
\beq
     \vDd{\sigma}{\eta}
  =  \bbigg{\sigmaSI \FSIQ + \sigmaSD \FSDQ}
     \sin(2 \eta)
\~,
\label{eqn:dsigma_deta}
\eeq
 and
\beq
     Q(\vchiLab, \eta) \vDd{\sigma}{\eta}
  =  2 \bbrac{\afrac{2 \mrN^2}{\mN} \vchiLab^2}
     \bbigg{\sigmaSI \FSIQ + \sigmaSD \FSDQ}
     \cos^3(\eta)
     \sin(\eta)
\~.
\label{eqn:QQ-dsigma_deta}
\eeq
 Hence,
 ignoring the nuclear form factor suppression,
 one can obtain the lower bounds of
 the most frequent and the most energetic recoil angles of
 $\eta_{\rm bound} = 45^{\circ}$
 and $\eta_{\rm {\it Q}, bound} = 30^{\circ}$,
 respectively,
 at which
 $|d\sigma / d\eta|$ and $Q(\eta) |d\sigma / d\eta|$
 are maximal
 \cite{DMDDD-NR}.

\begin{figure} [t!]
\begin{center}
 \begin{subfigure} [c] {5.5 cm}%
  \OnlinePlotNRchieta
   {NR}
   {20}
   {\includegraphics [width = 5.5 cm]
     {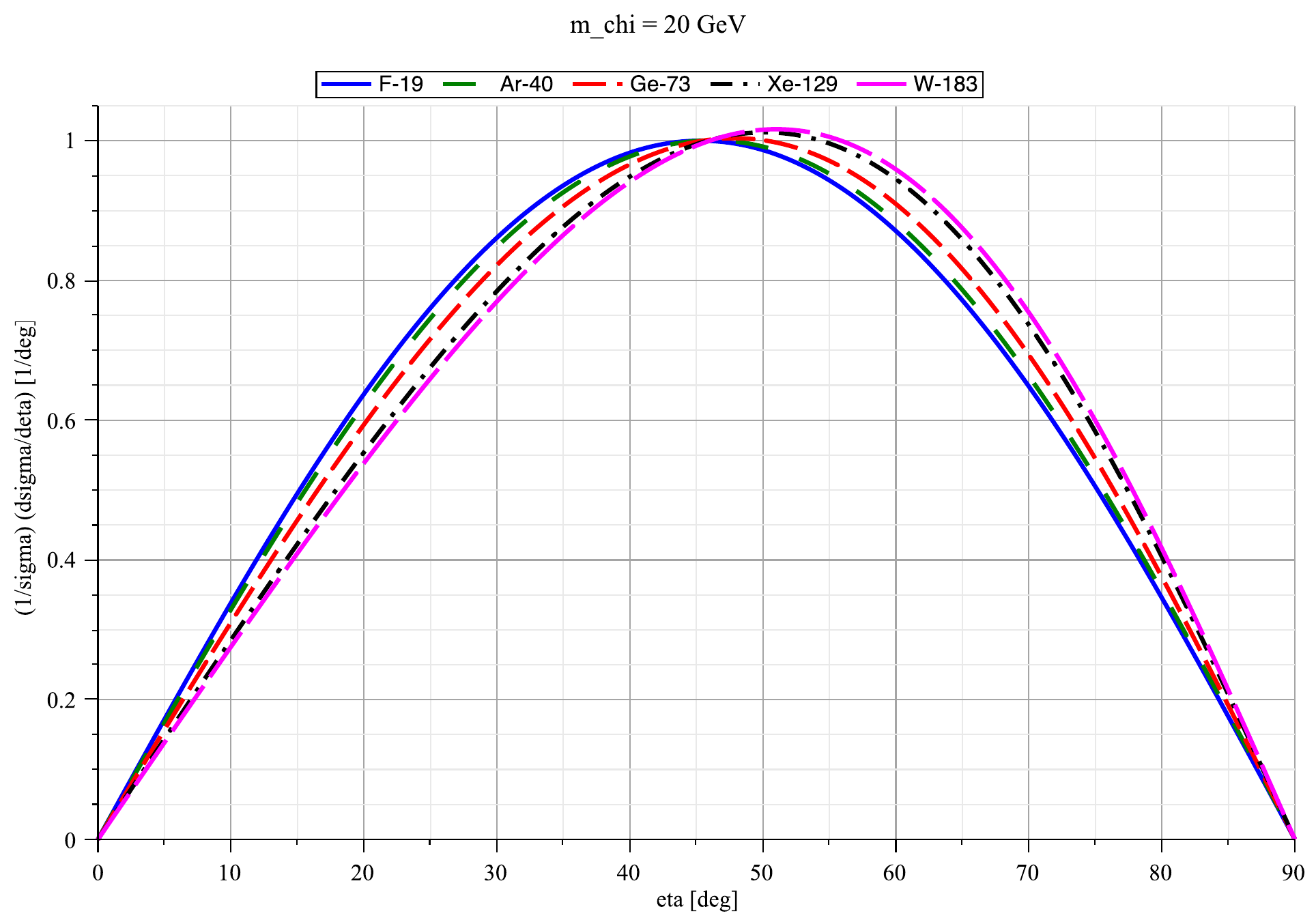}}%
 \caption{$\mchi =  20$ GeV}
 \end{subfigure}
 \begin{subfigure} [c] {5.5 cm}%
  \OnlinePlotNRchieta
   {NR}
   {100}
   {\includegraphics [width = 5.5 cm]
     {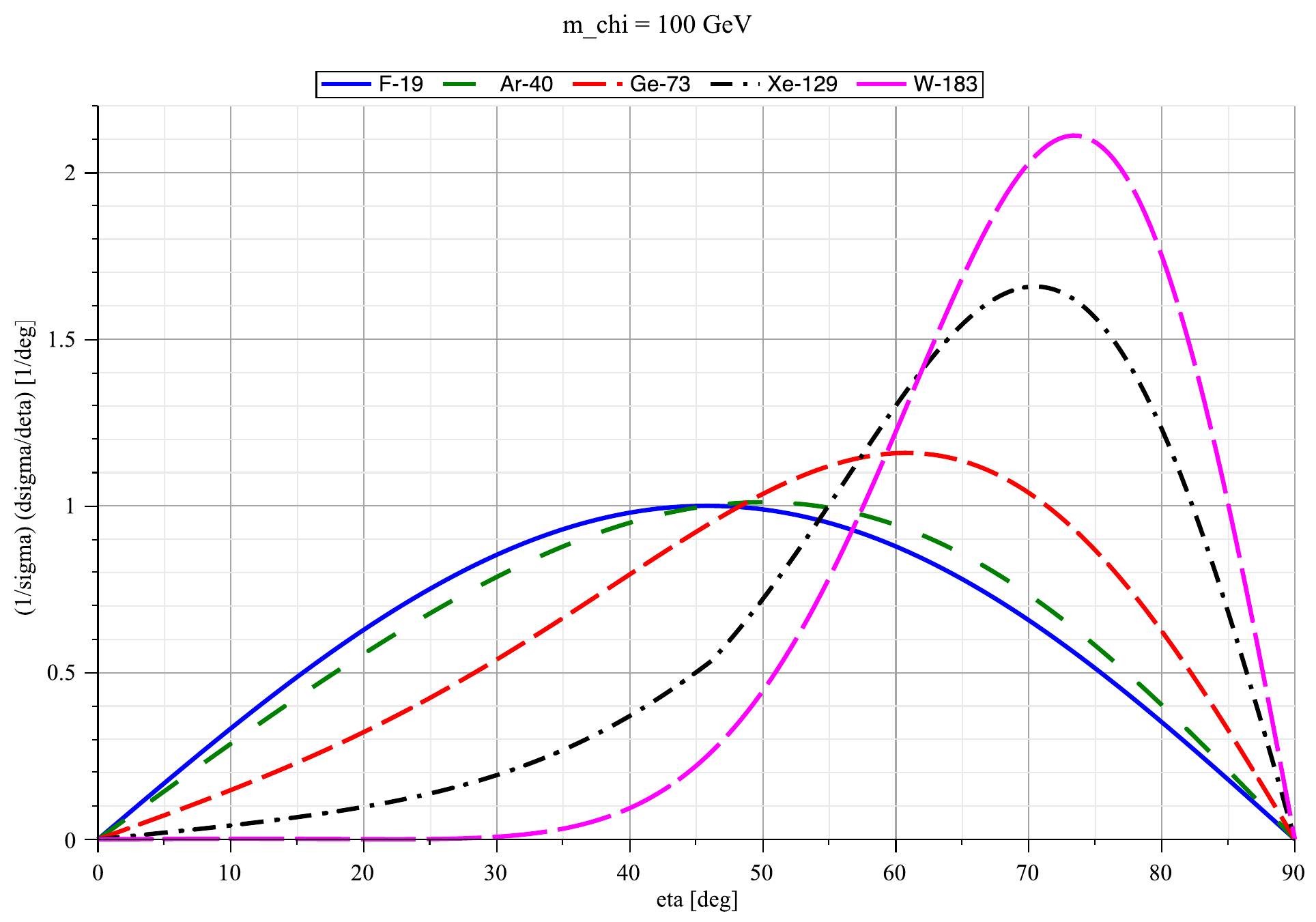}}%
 \caption{$\mchi = 100$ GeV}
 \end{subfigure}
 \begin{subfigure} [c] {5.5 cm}%
  \OnlinePlotNRchieta
   {NR}
   {500}
   {\includegraphics [width = 5.5 cm]
     {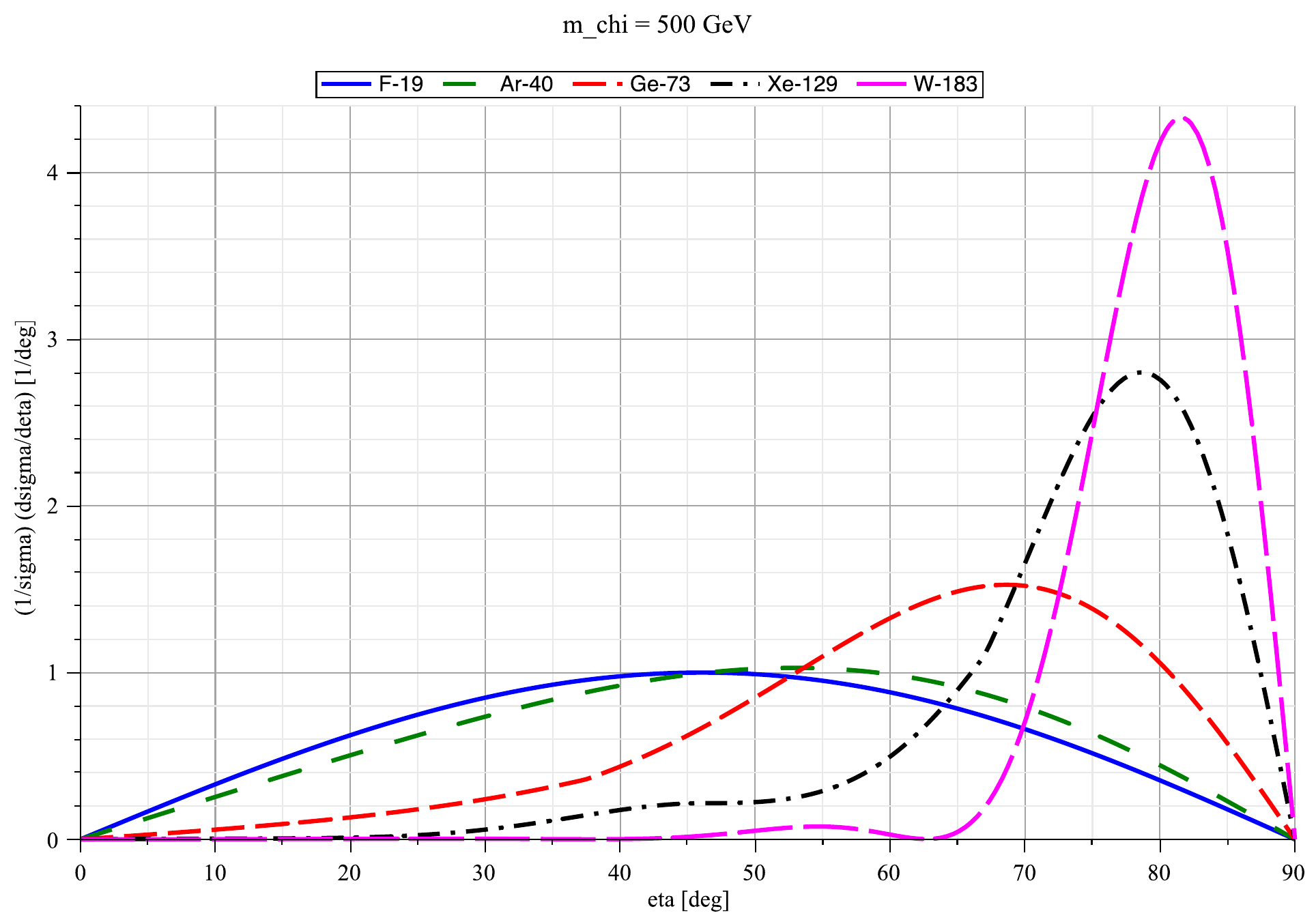}}%
 \caption{$\mchi = 500$ GeV}
 \end{subfigure}
\end{center}
\caption{
 The normalized recoil--angle dependence of
 $|d\sigma / d\eta|$
 given in Eq.~(\ref{eqn:dsigma_deta})
 for
 five frequently used target nuclei:
 $\rmF$     (solid        blue),
 $\rmAr$    (rare--dashed green),
 $\rmGe$    (dashed       red),
 $\rmXe$    (dash--dotted black),
 and $\rmW$ (long--dashed magenta).
 Three different WIMP masses
 have been considered.
 The form factors corresponding to
 the SI and the SD cross sections
 adopted in our simulation package
 have been used
 \cite{DMDDD-3D-WIMP-N}.
 The WIMP incident velocity here
 has been assumed (monotonically)
 as the root--mean--square velocity
 of halo WIMPs
 estimated by Eq.~(40) of Ref.~\cite{DMDDD-3D-WIMP-N}.
}
\label{fig:maple-dsigma_deta-F-Ar-Ge-Xe-W}
\end{figure}
\begin{figure} [b!]
\begin{center}
 \begin{subfigure} [c] {5.5 cm}%
  \OnlinePlotNRchieta
   {Q}
   {20}
   {\includegraphics [width = 5.5 cm]
     {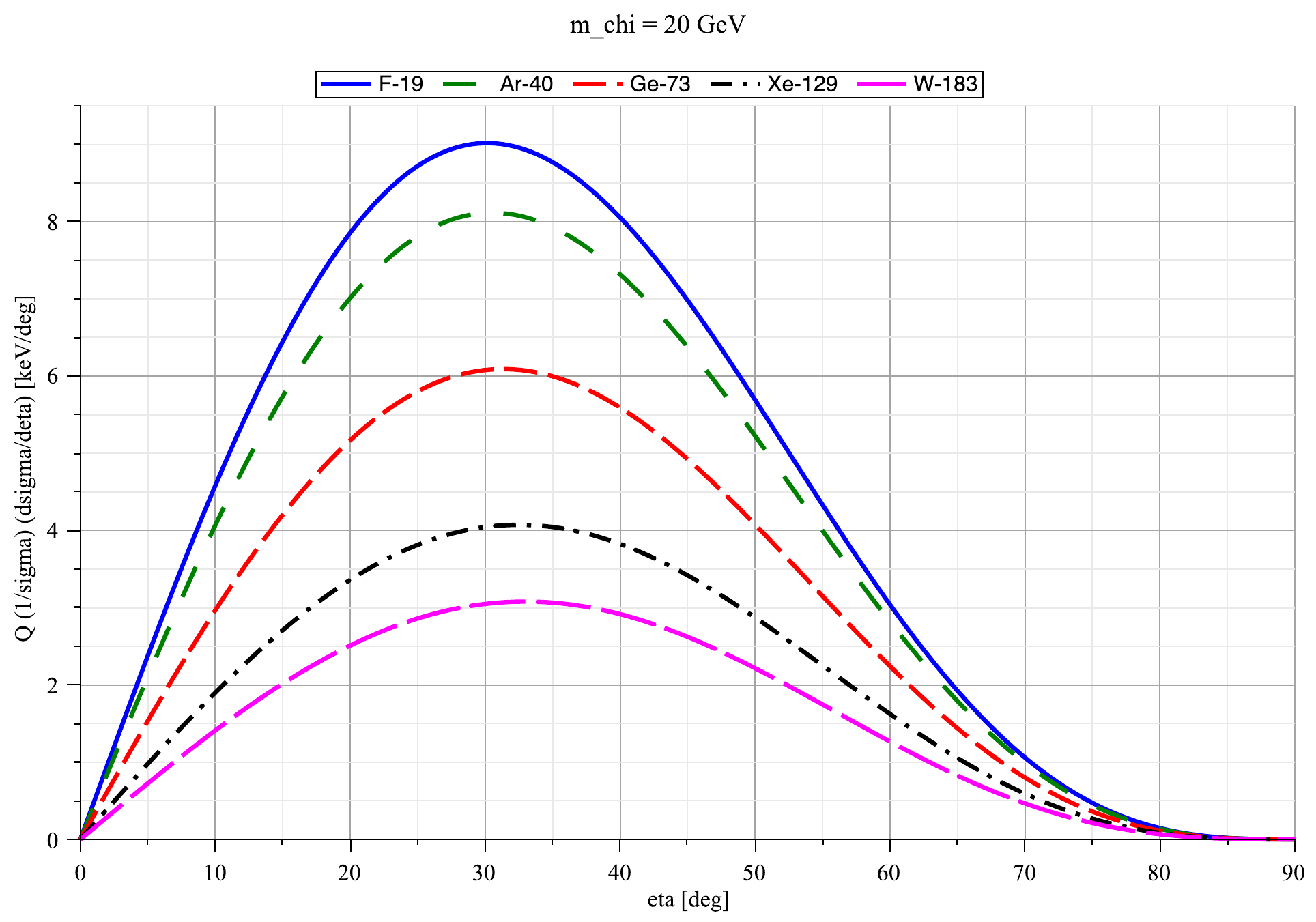}}%
 \caption{$\mchi =  20$ GeV}
 \end{subfigure}
 \begin{subfigure} [c] {5.5 cm}%
  \OnlinePlotNRchieta
   {Q}
   {100}
   {\includegraphics [width = 5.5 cm]
     {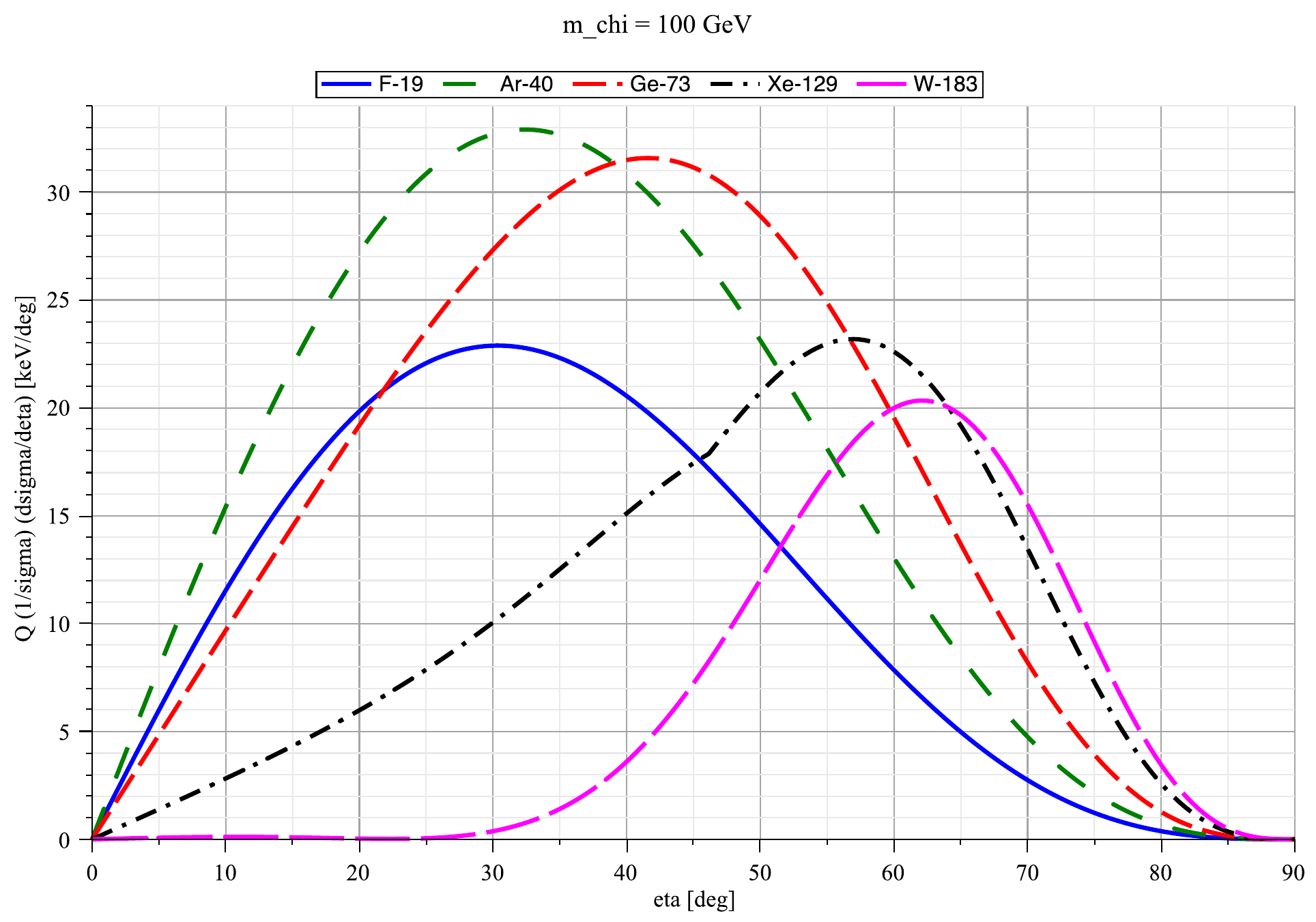}}%
 \caption{$\mchi = 100$ GeV}
 \end{subfigure}
 \begin{subfigure} [c] {5.5 cm}%
  \OnlinePlotNRchieta
   {Q}
   {500}
   {\includegraphics [width = 5.5 cm]
     {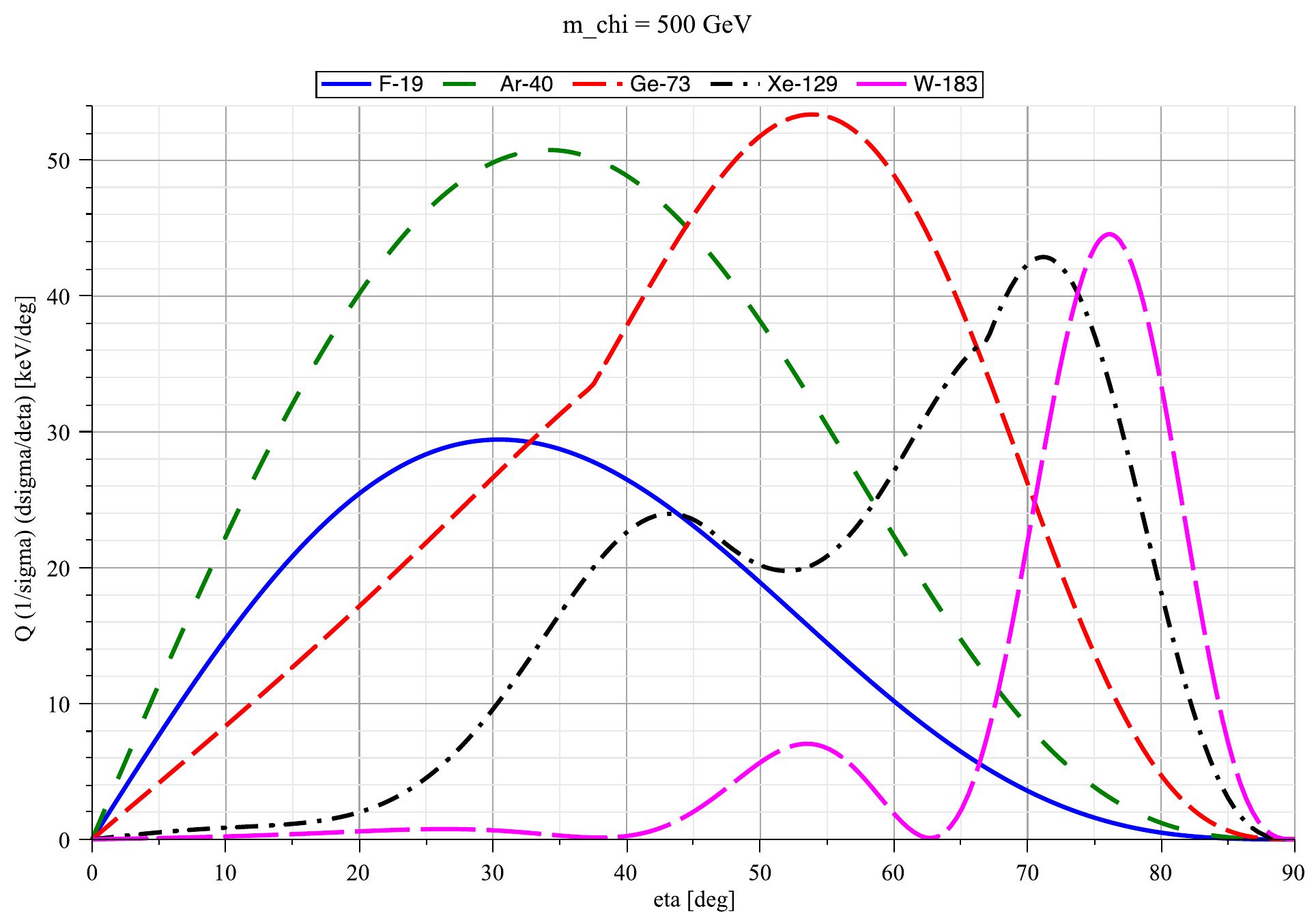}}%
 \caption{$\mchi = 500$ GeV}
 \end{subfigure}
\end{center}
\caption{
 Corresponding to Figs.~\ref{fig:maple-dsigma_deta-F-Ar-Ge-Xe-W}:
 the recoil--angle dependence of
 $Q(\eta) |d\sigma / d\eta|$
 given in Eq.~(\ref{eqn:QQ-dsigma_deta})
 for
 five frequently used target nuclei
 with three different WIMP masses.
 Notations as in Figs.~\ref{fig:maple-dsigma_deta-F-Ar-Ge-Xe-W}.
}
\label{fig:maple-Q-dsigma_deta-F-Ar-Ge-Xe-W}
\end{figure}

 In Figs.~\ref{fig:maple-dsigma_deta-F-Ar-Ge-Xe-W}
 and \ref{fig:maple-Q-dsigma_deta-F-Ar-Ge-Xe-W},
 we show
 the normalized recoil--angle dependence of
 $|d\sigma / d\eta|$
 and the corresponding recoil--angle dependence of
 $Q(\eta) |d\sigma / d\eta|$
 given in Eqs.~(\ref{eqn:dsigma_deta})
 and (\ref{eqn:QQ-dsigma_deta}),
 respectively.
 Five frequently used target nuclei:
 $\rmF$     (solid        blue),
 $\rmAr$    (rare--dashed green),
 $\rmGe$    (dashed       red),
 $\rmXe$    (dash--dotted black),
 and $\rmW$ (long--dashed magenta)
 and three different WIMP masses:
 $\mchi = 20$, 100, and 500 GeV
 have been considered%
\footnote{
 Interested readers can click each plot
 in Figs.~\ref{fig:maple-dsigma_deta-F-Ar-Ge-Xe-W}
 and \ref{fig:maple-Q-dsigma_deta-F-Ar-Ge-Xe-W}
 to open the corresponding webpage of
 the animated demonstration of
 detailed Monte Carlo simulation results
 with varying target nuclei
 (for more considered WIMP masses)
 (see Ref.~\cite{DMDDD-NR}
  for detailed discussions).
 Note however that
 all recoil--angle--dependence plots
 shown in our demonstration webpage
 \cite{AMIDAS-2D-web}
 are given as functions of
 the equivalent recoil angle $\thetaNRchi = \pi / 2 - \eta$.
}.
 The form factors corresponding to
 the SI and the SD cross sections
 adopted in our simulation package
 have been used
 \cite{DMDDD-3D-WIMP-N}.
 Note that
 these curves are WIMP--velocity dependent
 and
 the root--mean--square velocity
 of halo WIMPs
 estimated by Eq.~(40) of Ref.~\cite{DMDDD-3D-WIMP-N}
 has been considered again here.

 In these plots,
 the target and WIMP--mass dependent shifts of
 the most frequent and the most energetic recoil angles
 towards larger $\eta$'s
 caused by the nuclear form factor suppression
 can be seen obviously.
 This implies that,
 for different target nuclei,
 even with the same WIMP incident velocity,
 the scattering probability
 at each incident velocity--recoil angle ($\vchiLab$--$\eta$) combination
 along the intersection paths of
 the equal--recoil--energy planes
 shown in Fig.~\ref{fig:Qeta-v_l-v_h-Q_l-Q_h}(a)
 and thus
 the contributions
 to the differential event rate $dR / dQ$
 would be different
 \cite{DMDDD-v_theta}.

 Finally,
 it would be important to argue that,
 from Eq.~(\ref{eqn:dsigma_deta}),
 one can also find that
\beq
     \vDd{\sigma}{\Omega}
  =  \Dd{\eta}{\Omega}
     \vDd{\sigma}{\eta}
  =  \frac{1}{2 \pi \sin(\eta)}
     \vDd{\sigma}{\eta}
  = \frac{1}{\pi}
     \bbigg{\sigmaSI \FSIQ + \sigmaSD \FSDQ}
     \cos(\eta)
\~.
\label{eqn:dsigma_dOmega}
\eeq
 It seems thus to imply that
 the WIMP--induced nuclear recoil flux has a maximum
 at the zero--recoil--angle direction.
 However,
 it is actually just because that
 $d\Omega = 2 \pi \sin(\eta) d\eta$
 is infinitesimal
 at $\eta = 0$.

 It would also be worth to remind here that
 what we have discussed in this subsection indicates
 the importance of
 the form of the interaction(s)
 between incident WIMPs (Dark Matter particles) and
 the target nuclei (scattered ordinary particles)
 in the estimate of
 the scattering event rate
 (and,
  consequently,
  the determination of
  the exclusion limit
  or WIMP/DM properties),
 in particular,
 the relation between
 the recoil energy and
 the recoil angle
 (i.e.,
  the kinematics of WIMP/DM scattering).
 Hence,
 in other DM--interaction scenarios
 (e.g.~DM--electron scattering,
  cosmic--ray boosted DM,
  etc.),
 similar expressions
 for estimating the scattering event rates
 would also need to be revisited.

\subsection{1-D WIMP effective velocity distribution}
\label{sec:f1v_eff}

 By taking into account
 the flux proportionality
 to the WIMP incident velocity,
 the scattering probability distribution of
 incident halo WIMPs
 {\em with a given incoming velocity $\vchiLab$}
 off target nuclei
 going into a recoil angle of
 $\eta \pm d\eta / 2$
 in a recoil energy window of $Q \pm dQ / 2$
 can generally be expressed by
 \cite{DMDDD-3D-WIMP-N}
\beq
     f_{{\rm N_R}, \chiin}(\vchiLab, \eta)
  =  \frac{\vchiLab}{v_{\chi, {\rm cutoff}}}
     \vDd{\sigma}{\eta}
  =  \frac{\vchiLab}{v_{\chi, {\rm cutoff}}}
     \bbigg{\sigmaSI \FSIQ + \sigmaSD \FSDQ}
     \sin(2 \eta)
\~.
\label{eqn:f_NR_eta}
\eeq
 Then,
 as argued in Sec.~\ref{sec:FQ}
 and demonstrated in Ref.~\cite{DMDDD-fv_eff},
 due to the nuclear form factor suppression
 appearing as the second (bracket) factor
 in Eq.~(\ref{eqn:f_NR_eta}),
 the actual ``effective'' velocity distribution of
 incident halo WIMPs
 {\em scattering off} target nuclei
 should be (much) more strongly reduced
 from the theoretical prediction
 in the high velocity range
 than in the low velocity range,
 once the target nuclei and incident WIMPs are heavy.
 This reduction
 could however be balanced by
 the (first) factor of
 the flux proportionality to the WIMP incident velocity
 in Eq.~(\ref{eqn:f_NR_eta}).
 Consequently,
 the ``effective'' average and root--mean--square velocities of
 light (heavy) WIMPs (impinging on and) scattering off light (heavy)  target nuclei
 should be larger (smaller) than
 those of the entire Galactic halo WIMPs
 \cite{DMDDD-fv_eff}.

 However,
 in a standard derivation of
 the differential event rate (\ref{eqn:dRdQ_SISD})
 given in,
 e.g.~Ref.~\cite{SUSYDM96},
 the average velocity of incident halo WIMPs:
\beq
     \expv{v}
  =  \intz v f_1(v) \~ dv
\~,
\eeq
 has been adopted directly
 for estimating the flux of
 the incident and also the ``scattering'' WIMPs
 needed in $dR /dQ$.
 As argued above,
 for a given WIMP incident velocity
 but different recoil energies
 (on different equal--recoil--energy planes
  in Fig.~\ref{fig:Qeta-v_l-v_h-Q_l-Q_h}(a)),
 the scattering probability of
 different corresponding recoil angles
 and the contribution to the differential event rate
 should be different.
 This indicates that,
 even if one consider to use an average velocity
 as a macroscopical description/approximation for
 incident velocities of the ``scattering'' WIMPs,
 a modified effective velocity distribution
 depending on the considered recoil energy
 (as well as
  on the target nucleus and the WIMP mass)
 would be needed
 (see Ref.~\cite{DMDDD-v_theta}
  for detailed numerical simulation results).

 Finally,
 it would be worth to note that
 the scattering probability given in Eq.~(\ref{eqn:f_NR_eta})
 (including the nuclear form factor
  in the bracket
  and the recoil angle term of $\sin(2 \eta)$)
 would need to be included inside the integral
 over the velocity distribution function
 in Eq.~(\ref{eqn:dRdQ_SISD})
 and the integral would need to cover
 all possible incident velocity--recoil angle combination.

%
%

%
 %
%
\section{Double differential event rate}
\label{sec:d2RdQdOmega}

 In this section,
 we discuss then
 the double differential event rate
 used in directional direct DM detection physics.

 In literature,
 the expression for
 the double differential event rate
 for elastic WIMP--nucleus scattering
 has been given by
 \cite{Gondolo02,
       Billard10d,
       OHare14, Mayet16,
       OHare17,
       Vahsen20}
\beqn
     \frac{d^2 R}{dQ \~ d\Omega_{\hbf{q}}}
 \=  \frac{\rho_0}{4 \pi \mchi \mrN^2}
     \bbigg{\sigmaSI \FSIQ + \sigmaSD \FSDQ}
     \non\\
 \conti ~~~~ ~~~~ ~~ \times 
     \intvminQ
     \delta\abrac{\VchiLab \cdot \hbf{q} - \frac{q}{2 \mrN}}
     f(\VchiLab) \~ d^3\VchiLab
\~.
\label{eqn:d2RdQdOmega_SISD}
\eeqn
 Here
 $\Omega_{\hbf{q}}$ is the solid angle
 around the recoil direction $\hbf{q}$
 with the recoil angle $\eta$
 and the $\delta$--function ensures that,
 for an incident WIMP
 moving with a given velocity $\vchiLab \ge \vmin(Q)$
 and transferring energy $Q$ to
 a nucleus in our detector,
 the recoil angle of
 the scattered target nucleus $\eta$ satisfies that
\beq
     \VchiLab \cdot \hbf{q}
  =  \vchiLab \cos\eta
  =  \frac{q}{2 \mrN}
  =  \sfrac{\mN}{2 \mrN^2} \~ \sqrt{Q}
\~.
\label{eqn:V_dot_q}
\eeq

 For the use of
 the double differential event rate
 given in Eq.~(\ref{eqn:d2RdQdOmega_SISD}),
 it seems that
 one not only assumes implicitly {\em again} that
 the velocity distribution of
 the ``scattering'' WIMPs
 (moving with velocities larger than $\vmin(Q)$)
 obeys exactly
 the 1-D velocity distribution of
 ``incident'' halo WIMPs
 in the Equatorial/laboratory frames,
 but also {\em ignores
 the recoil--angle dependence of
 the differential scattering cross section}
 given in Eq.~(\ref{eqn:dsigma_eta}).
 Then,
 for the considered recoil energy $Q \pm dQ / 2$,
 the angular distribution of the recoil flux
 is determined by Eq.~(\ref{eqn:V_dot_q})
 with incident velocities
 constrained by the WIMP velocity distribution.

 However,
 ...

\subsection{(Only) the relation between
            the WIMP incident and the nuclear recoil directions}
\label{sec:V_dot_q}

 The double differential event rate (\ref{eqn:d2RdQdOmega_SISD})
 seems to describe only the relation between
 the WIMP incident and the nuclear recoil directions.
 Namely,
 considering at first the case that
 all incident WIMPs move (almost) identically
 in one common direction,
 only the recoil angles of
 the scattering events matter
 and
 the angular distribution of the recoil flux
 would be displayed
 with the azimuthal symmetry
 around this common direction.

 However,
 once we consider two bunches of WIMPs
 moving in two different common directions,
 the situation seems to become unsolvable.
 Giving an energy window of $Q \pm dQ / 2$,
 according to Eq.~(\ref{eqn:d2RdQdOmega_SISD}),
 the first bunch of WIMPs
 moving with different velocities
 would induce nuclear recoils in different rings
 centered at the first common direction of the incident velocities;
 so the second bunch of WIMPs
 and the induced nuclear recoils.
 The question now is:
 how could we combine these two recoil distribution patterns?
 And,
 with the third,
 the forth recoil distribution patterns?

 Unfortunately,
 ...

\subsection{WIMP incident directions would not be (highly) concentrated}
\label{sec:WIMP_wind}

 In literature,
 (for practical uses of
  the double differential event rate
  (\ref{eqn:d2RdQdOmega_SISD}),)
 it is often to consider
 the opposite direction of
 the movement of the Solar system
 in our Galaxy,
 namely,
 the direction
 from the Cygnus constellation
 to the Solar center,
 as a reference direction
 and
 assume that
 the incident velocities of halo WIMPs
 highly concentrate
 around this direction of the so--called ``WIMP wind''
 \cite{Billard10d, Mayet16,
       Vahsen20, Vahsen21}.

 However,
 the real situation should more likely be that
 described by R.~J.~Creswick {\it et al.}~%
 \cite{Creswick10}:
\begin{center}
\begin{minipage}{15.75 cm}
 ``{\it
    Several effects mitigate against a strong correlation
    between the distribution of recoils and the direction of the WIMP wind.
    First,
    the velocity {\boldmath$v$} of the WIMP
    is the vector sum of the fixed wind {\boldmath$w$}
    and a random virial velocity, {\boldmath$u$},
    of roughly equal magnitude.
    Hence
    at any given time
    the direction of the WIMP impinging upon the crystal
    has a broad distribution
    that is only peaked around {\boldmath$w$}.
    The directionality effect is further diluted
    by the recoil of the elastically scattered nucleus,
    which we assume is isotropic in the CM frame\footnotemark.
    Finally,
    the directionality effect is further reduced
    by the nuclear form factor,
    relevant for heavier nuclei,
    which suppresses scattering a high momentum transfer.}''
\end{minipage}
\end{center}
\footnotetext{
 From this isotropy of the recoil angle
 in the ``center--of--mass'' frame
 \cite{Cooley21},
 one can derive directly
 the recoil--angle dependence (\ref{eqn:dsigma_eta}) of
 the differential cross section
 to the differential recoil angle
 in the ``laboratory'' frame.
}
 In the Galactic point of view,
 the moving directions of incident WIMPs
 should be random and (approximately) isotropic,
 then
 (vector) added by
 the movement of the Solar system
 (or,
  more precisely,
  that of the Earth or even the laboratory)
 to return the incoming directions
 in the laboratory frame.
 Hence,
 the WIMP incident flux should indeed center
 at (very closely to)
 the direction
 from the Cygnus constellation
 to the Solar center.
 However,
 its decreasing distribution
 would not be ring--like around the center
 but rather distorted.

\begin{figure} [t!]
\begin{center}
 \begin{subfigure} [c] {5.5 cm}
  \includegraphics [width = 5.5 cm]
   {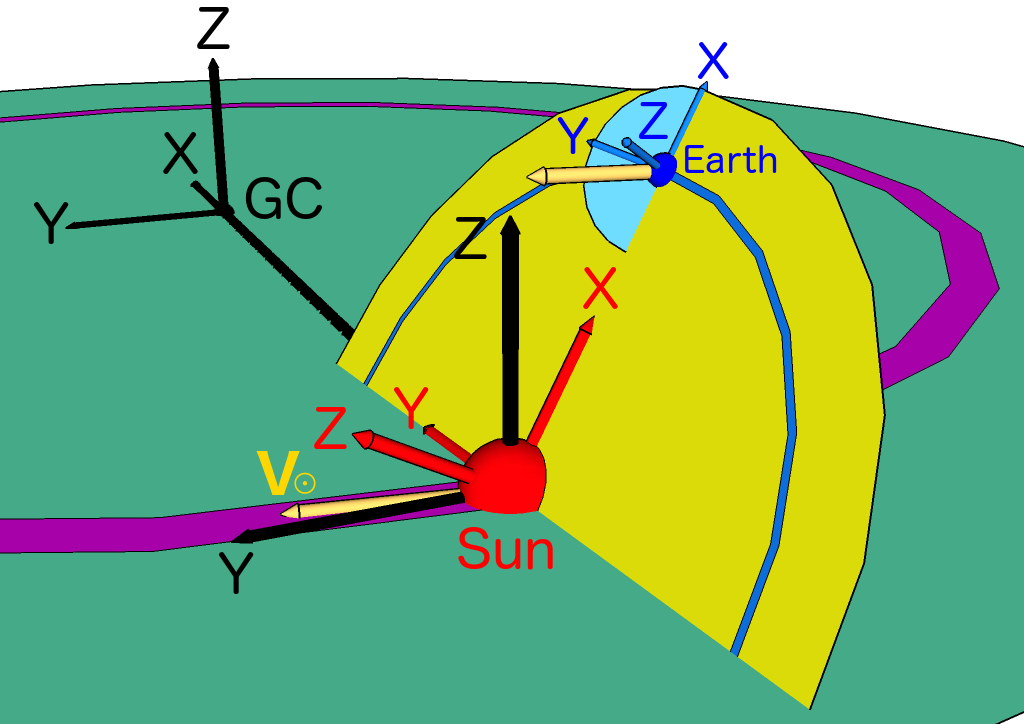}
 \caption{}
 \end{subfigure}
 \begin{subfigure} [c] {5.5 cm}
  \includegraphics [width = 5.5 cm]
   {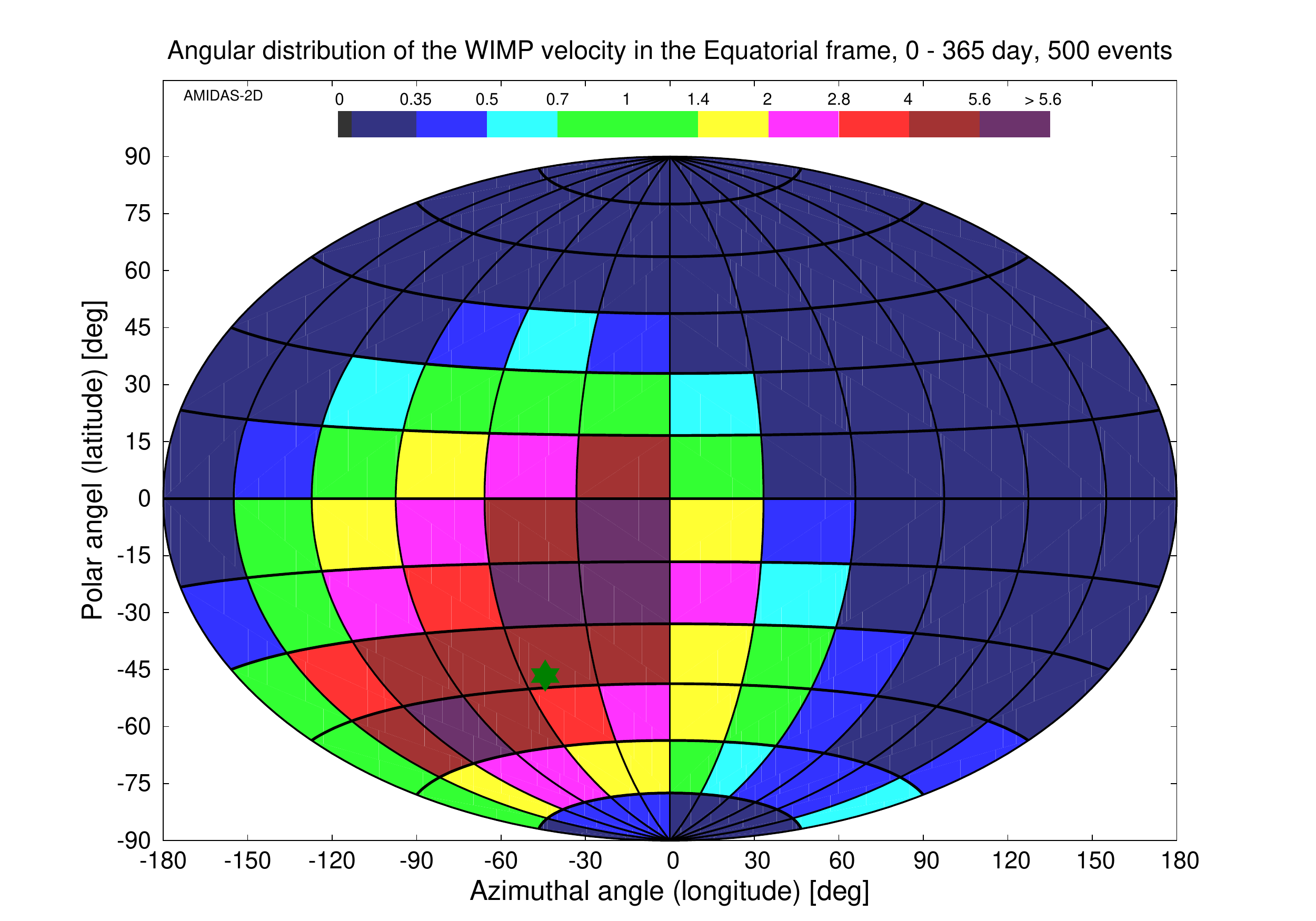}
 \caption{}
 \end{subfigure}
 \begin{subfigure} [c] {5.5 cm}
  \includegraphics [width = 5.5 cm]
   {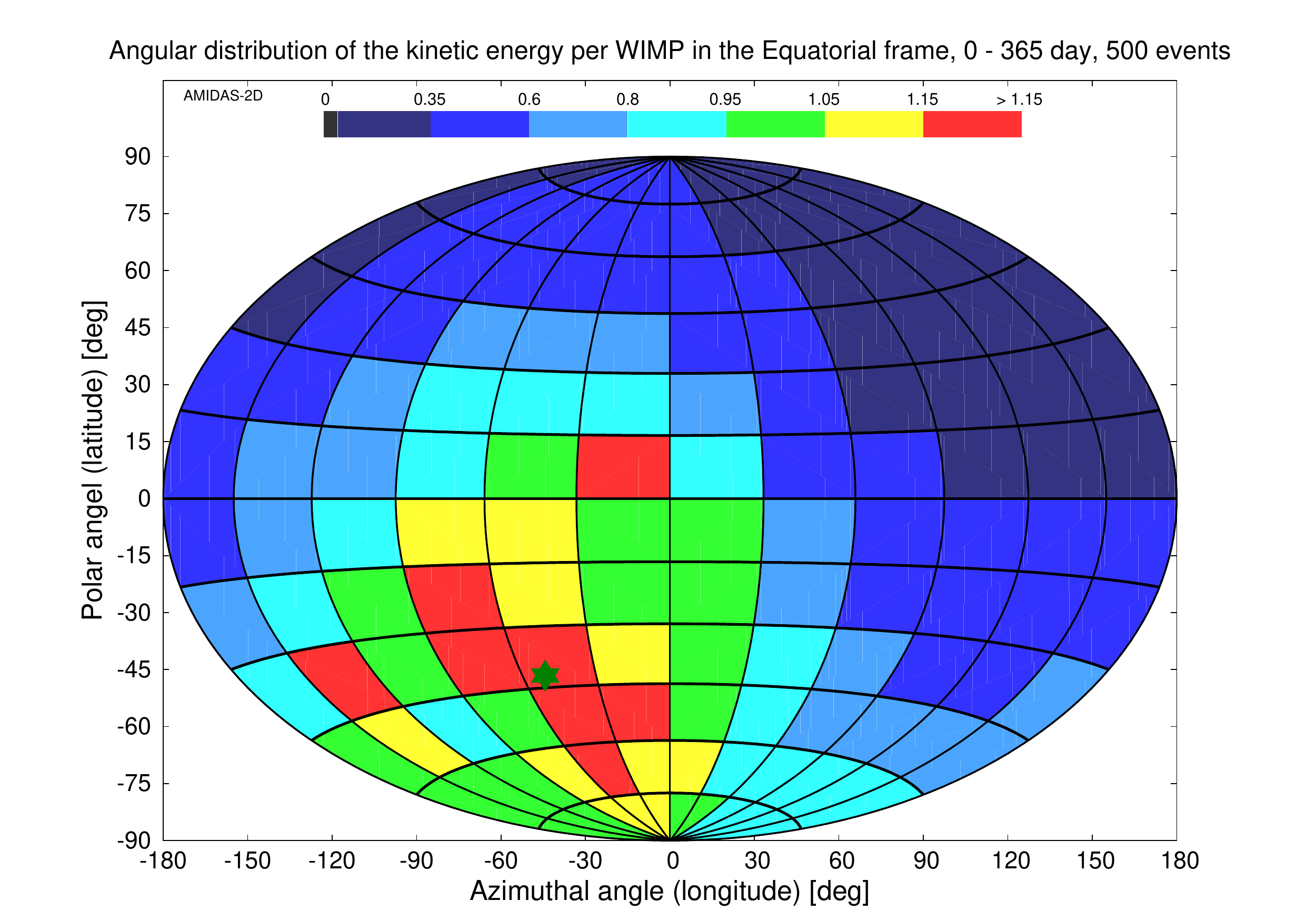}
 \caption{}
 \end{subfigure}
\end{center}
\caption{
 (a)
 The movement of the Solar system
 around the Galactic center (GC)
 (along the magenta circular path),
 which points currently
 towards the Cygnus constellation
 (indicated by the golden arrows).
 The relative orientations between
 the black Galactic,
 the red Ecliptic,
 and the blue Equatorial coordinate systems
 (on the date of the vernal equinox)
 have also been sketched,
 whereas
 the blue circular band
 (on the yellow Ecliptic plane)
 indicates
 the Earth's orbit around the Sun.
 (Figure from Ref.~\cite{DMDDD-N}).
 (b)(c)
 The simulated angular distributions of
 the WIMP incident flux
 and the ``average WIMP kinetic energy''
 (in unit of the all--sky average values)
 in the Equatorial coordinate system,
 respectively.
 500 simulated WIMP velocities on average
 (Poisson distributed)
 in one entire year
 have been recorded
 and binned into 12 $\times$ 12 bins
 for the azimuthal angle and the elevation,
 respectively.
 The dark--green stars
 indicate
 the opposite direction of the Solar Galactic movement
 \cite{Bandyopadhyay10}:
 42.00$^{\circ}$S, 50.70$^{\circ}$W.
 Note that
 the scales of the color bars
 are different.
 (Figures from Refs.~\cite{DMDDD-N} and \cite{DMDDD-P},
  respectively).
}
\label{fig:v_Sun_Eq-S-G-rotated}
\end{figure}

 In Fig.~\ref{fig:v_Sun_Eq-S-G-rotated}(a),
 the golden arrows
 indicate
 the movement of the Solar system
 around the Galactic center
 (along the magenta circular path),
 which points currently
 towards the Cygnus constellation.
 We show also
 the relative orientations between
 the (black) Galactic,
 the (red) Ecliptic,
 and the (blue) Equatorial coordinate systems
 (on the date of the vernal equinox)
 \cite{DMDDD-N}.
 Then
 we simulated 500 {\em isotropic} WIMP velocities on average
 (Poisson distributed)
 in the Galactic coordinate system
 in one entire year
 and transformed them
 to the Equatorial coordinate system
 \cite{DMDDD-N, DMDDD-3D-WIMP-N}.
 Their angular distribution
 is shown in Fig.~\ref{fig:v_Sun_Eq-S-G-rotated}(b)
 (in unit of the all--sky average value).
 The dark--green star
 indicates
 the opposite direction of the Solar Galactic movement
 in the Equatorial coordinate system
 \cite{Bandyopadhyay10}:
 42.00$^{\circ}$S, 50.70$^{\circ}$W.
 It can be found interestingly that,
 while
 the intense area (directions) of the WIMP incident flux
 covers indeed
 the opposite direction of the Solar Galactic movement,
 it distributes rather {\em obliquely}
 from the center
 towards the southwest corner of the sky
 \cite{DMDDD-N}.

 On the other hand,
 the simulated angular distribution of
 the ``average kinetic energy'' of
 incident WIMPs
 (in unit of the all--sky average value)
 in the Equatorial coordinate system
 in Fig.~\ref{fig:v_Sun_Eq-S-G-rotated}(c)
 shows that,
 firstly,
 in contrast to the obliquely--distributed intense area of
 the WIMP incident flux,
 the main hot spot of
 the average WIMP kinetic energy
 spreads longitudinally through
 the opposite direction of the Solar Galactic movement
 \cite{DMDDD-P}.
 Secondly and surprisingly,
 two extra peaks
 appear near the center
 and the southwest corner of the sky
 \cite{DMDDD-P}.

 These simulation results demonstrate clearly that,
 even though
 the WIMP Galactic velocity distribution
 has been here most simply assumed
 as isotropic,
 the angular distribution of
 the 3-D velocity of WIMPs
 impinging on our detectors
 would already be more complicated than
 a highly concentrated WIMP wind%
\footnote{
 As demonstrated in Ref.~\cite{DMDDD-fv_eff},
 the 3-D (radial and angular) distribution of
 the ``effective'' velocity distribution of WIMPs
 {\em scattering off} target nuclei
 in both of the Galactic and the Equatorial coordinate systems
 would be much more complicated
 and depend on target material and the WIMP mass.
}.

 Furthermore,
 ...

\subsection{WIMP--induced nuclear recoils
            could be strongly deflected from
            the WIMP--wind direction}
\label{sec:NR}

 Probably because that,
 with a given recoil energy $Q$,
 for calculating
 the minimal--required incoming velocity of incident WIMPs,
 $\vmin(Q)$,
 by Eq.~(\ref{eqn:QQ_eta}),
 the zero--recoil--angle ($\eta = 0$) condition has been used,
 or
 because that
 $d\sigma / d\Omega$ given in Eq.~(\ref{eqn:dsigma_dOmega})
 indicates that
 the differential WIMP--nucleus scattering cross section
 per unit solid angle
 has a maximum
 when $\eta = 0$,
 in literature,
 e.g.~Refs.~\cite{Billard09, Green10,
                  OHare14, OHare15b,
                  Riffard16, Mayet16, OHare17,
                  Baracchini20a,
                  Vahsen20, Vahsen21},
 it has very frequently been asserted that
 the most possible WIMP--induced events
 would be ``head--on'' scattering
 and thus
 the angular distribution of
 the recoil flux of
 the scattered target nuclei
 would peak opposite to the direction of
 the movement of the Solar system/Earth.
 Additionally,
 in e.g.~Refs.~\cite{Billard09,
                     OHare14,
                     Mayet16,
                     Vahsen20},
 the authors have even presented
 ring--like--decreased recoil--flux distributions
 centered at the direction of the Cygnus constellation/WIMP wind
 (estimated probably
  by using the double differential event rate
  (\ref{eqn:d2RdQdOmega_SISD})).

 However,
 as argued and demonstrated in several different ways
 in Sec.~\ref{sec:dRdQ},
 firstly,
 due to the {\em zero} differential cross section
 at zero recoil angle,
 the recoil direction of
 the scattered target nucleus
 will definitely not
 along the incident direction of
 the scattering WIMP.
 Instead,
 due to the lower bound and the increase of
 the most frequent recoil angle
 (with the increasing target and/or WIMP mass),
 the deviation of
 the nuclear recoil direction
 from that of the incident scattering WIMP
 could be pretty large.
 For heavy target nuclei like $\rmXe$ and $\rmW$
 (as well as
  $\rmXA{I}{127}$ and $\rmXA{Cs}{133}$),
 a large number of recoil events
 could even be deflected almost perpendicularly
 (see Figs.~\ref{fig:maple-dsigma_deta-F-Ar-Ge-Xe-W}(b) and (c)).

 Moreover,
 as demonstrated in Sec.~\ref{sec:WIMP_wind},
 the angular distribution of
 the WIMP incident flux
 in the Equatorial coordinate system
 should be more complicated than
 that of a highly concentrated WIMP--wind.
 This makes the ``superposition'' of
 the nuclear recoil fluxes
 induced by WIMPs
 coming (anisotropically) from all directions
 much more complicated
 than a simple ring--like--decreased distribution
 centered around the theoretical WIMP--wind direction.

\begin{figure} [t!]
\begin{center}
 \begin{subfigure} [c] {5.5 cm}%
  \OnlinePlotNRangEq
   {NR}
   {F19}
   {\includegraphics [width = 5.5 cm]
     {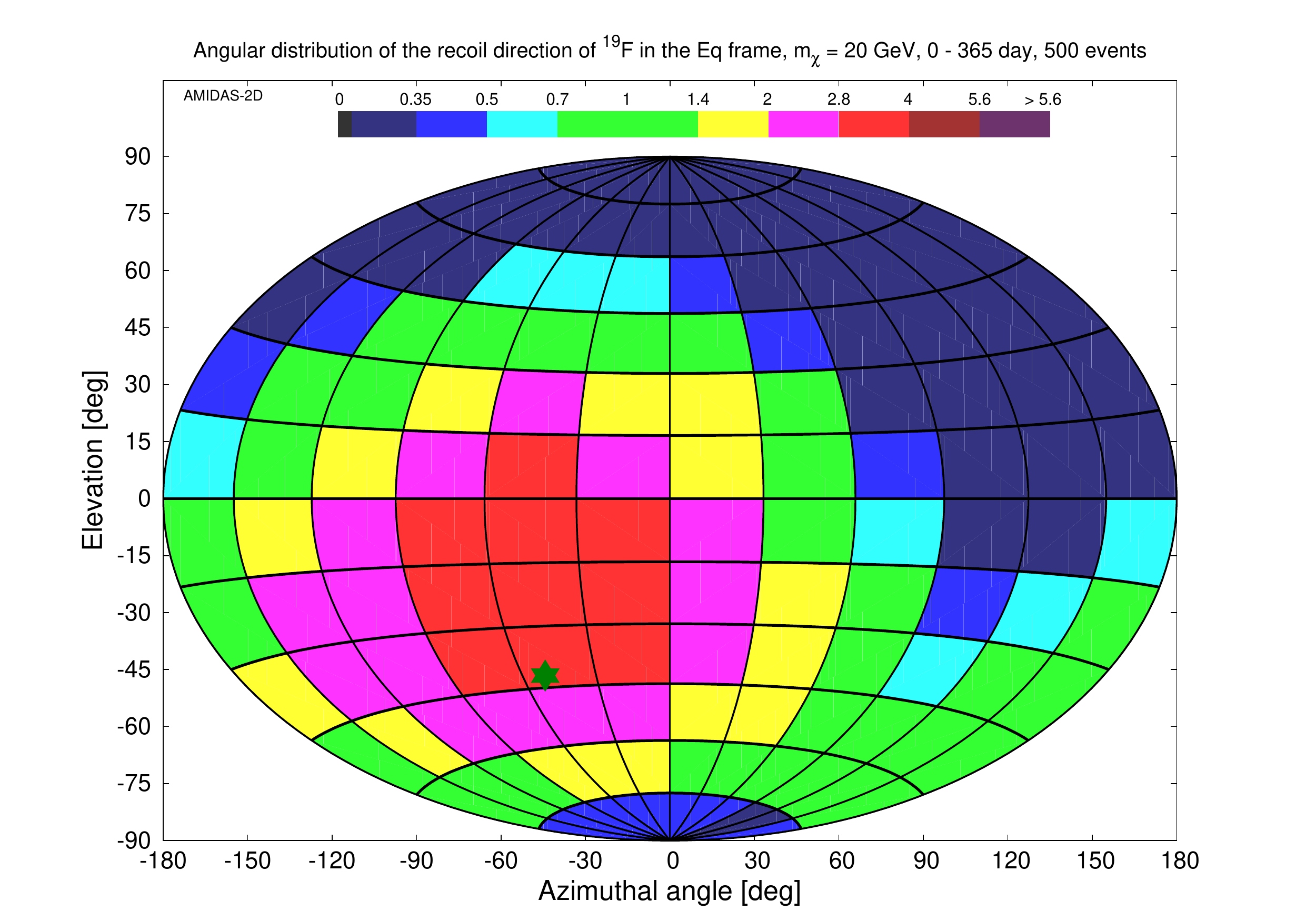}}%
 \caption{\footnotesize  20-GeV WIMPs off $\rmF$}
 \end{subfigure}
 \begin{subfigure} [c] {5.5 cm}%
  \OnlinePlotNRangEq
   {NR}
   {Ge73}
   {\includegraphics [width = 5.5 cm]
     {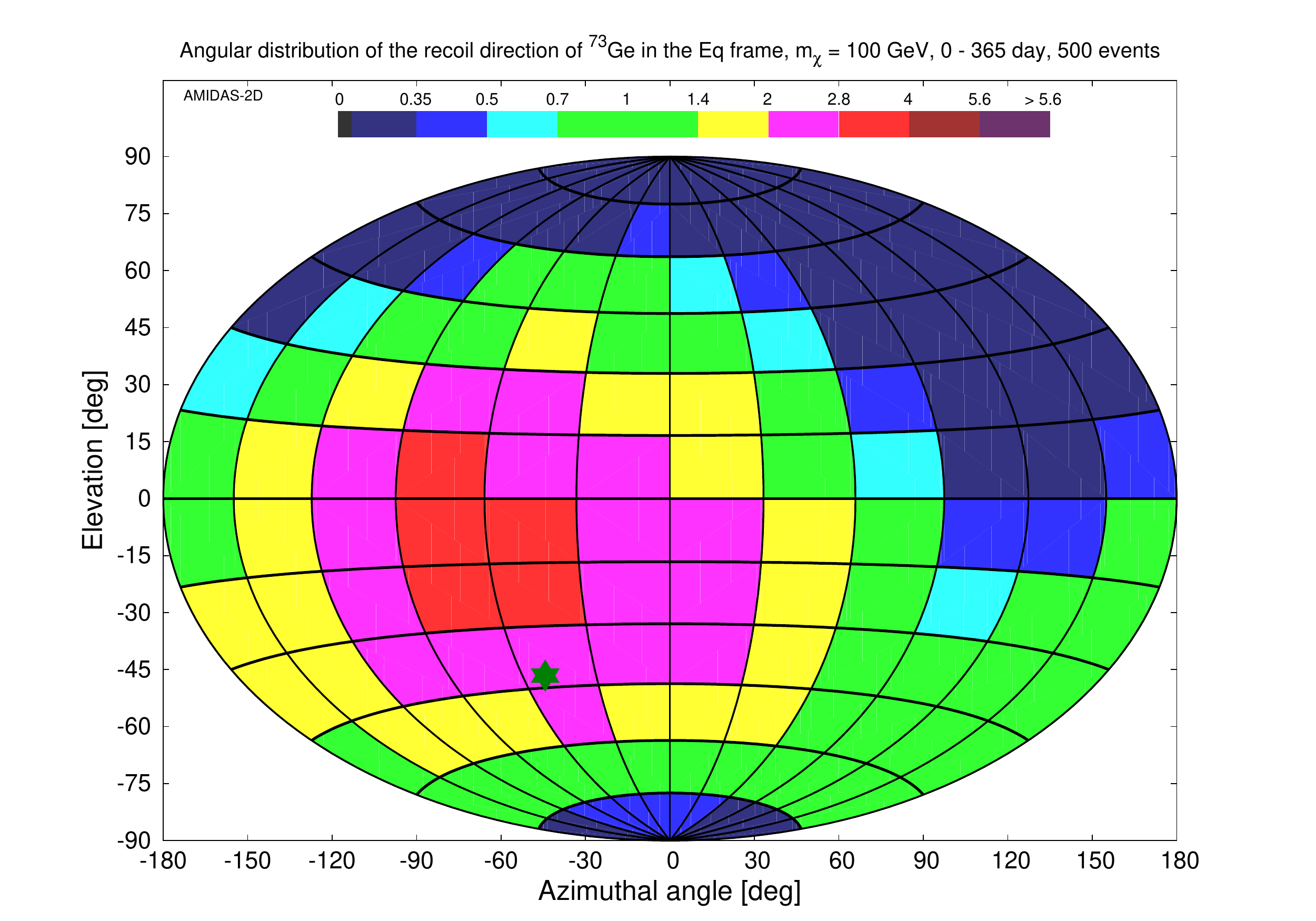}}%
 \caption{\footnotesize 100-GeV WIMPs off $\rmGe$}
 \end{subfigure}
 \begin{subfigure} [c] {5.5 cm}%
  \OnlinePlotNRangEq
   {NR}
   {Xe129}
   {\includegraphics [width = 5.5 cm]
     {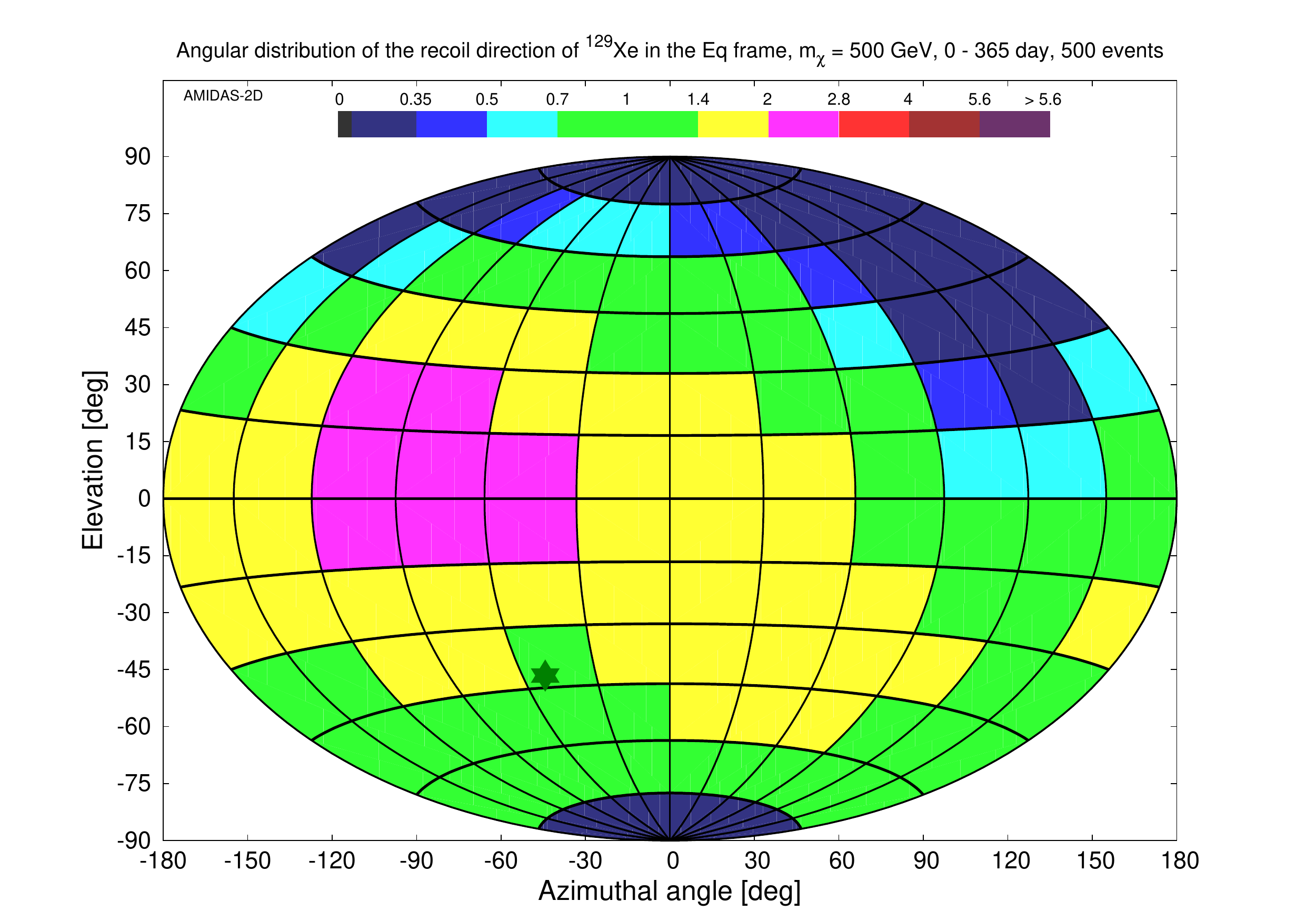}}%
 \caption{\footnotesize 500-GeV WIMPs off $\rmXe$}
 \end{subfigure}
\end{center}
\caption{
 The angular distributions of
 the WIMP--induced nuclear recoil flux
 observed in the Equatorial coordinate system
 (in unit of the all--sky average value).
 WIMPs of three different masses
 scatter off three different target nuclei
 have been presented.
 500 accepted
 scattering events on average
 (Poisson distributed)
 in one entire year
 have been recorded.
 (Figures from Refs.~\cite{DMDDD-NR, AMIDAS-2D-web}).
}
\label{fig:NR_ang-Eq-0500-00000}
\end{figure}
\begin{figure} [b!]
\begin{center}
 \begin{subfigure} [c] {5.5 cm}%
  \OnlinePlotNRangEq
   {QoN}
   {F19}
   {\includegraphics [width = 5.5 cm]
     {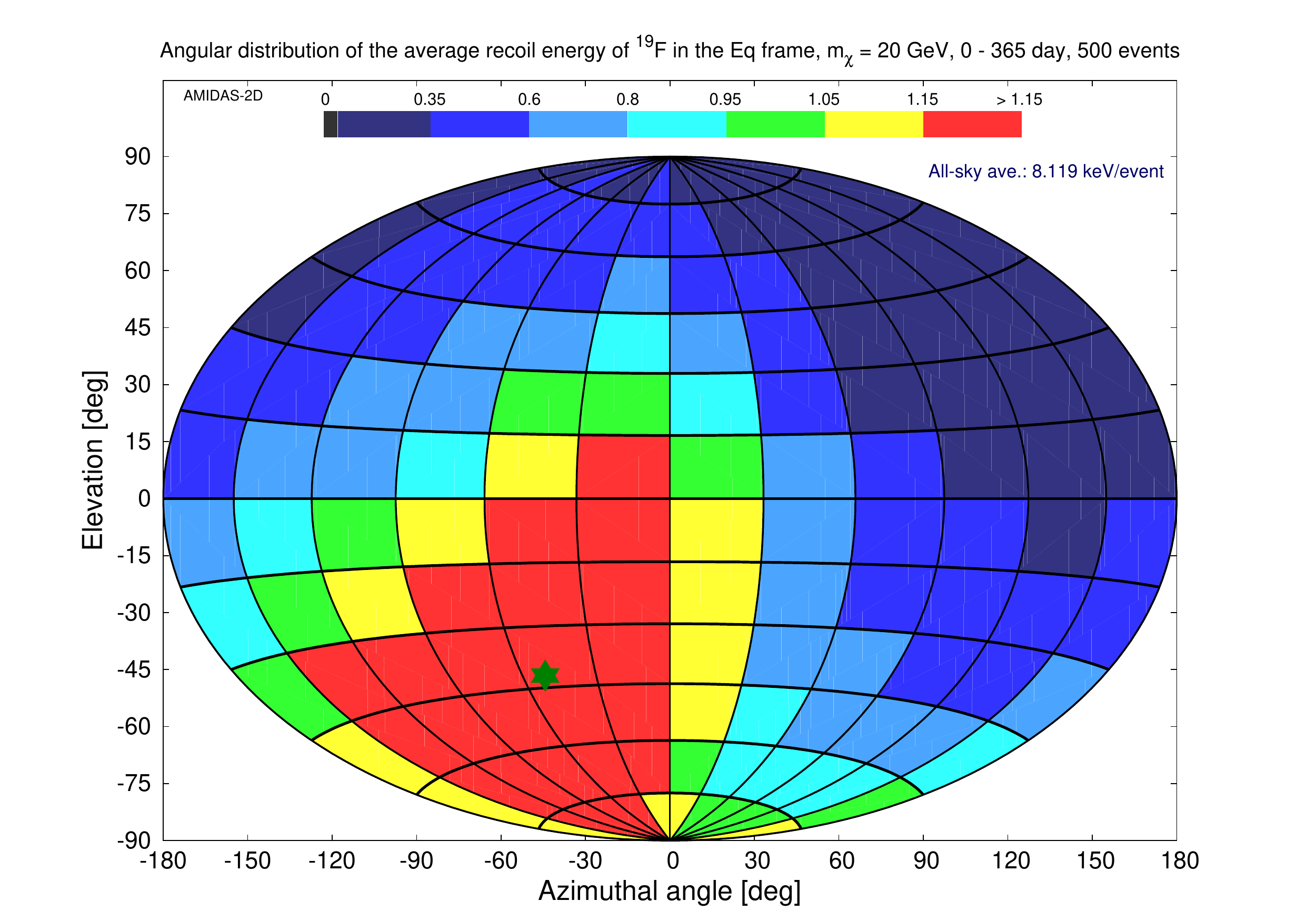}}%
 \caption{\footnotesize  20-GeV WIMPs off $\rmF$}
 \end{subfigure}
 \begin{subfigure} [c] {5.5 cm}%
  \OnlinePlotNRangEq
   {QoN}
   {Ge73}
   {\includegraphics [width = 5.5 cm]
     {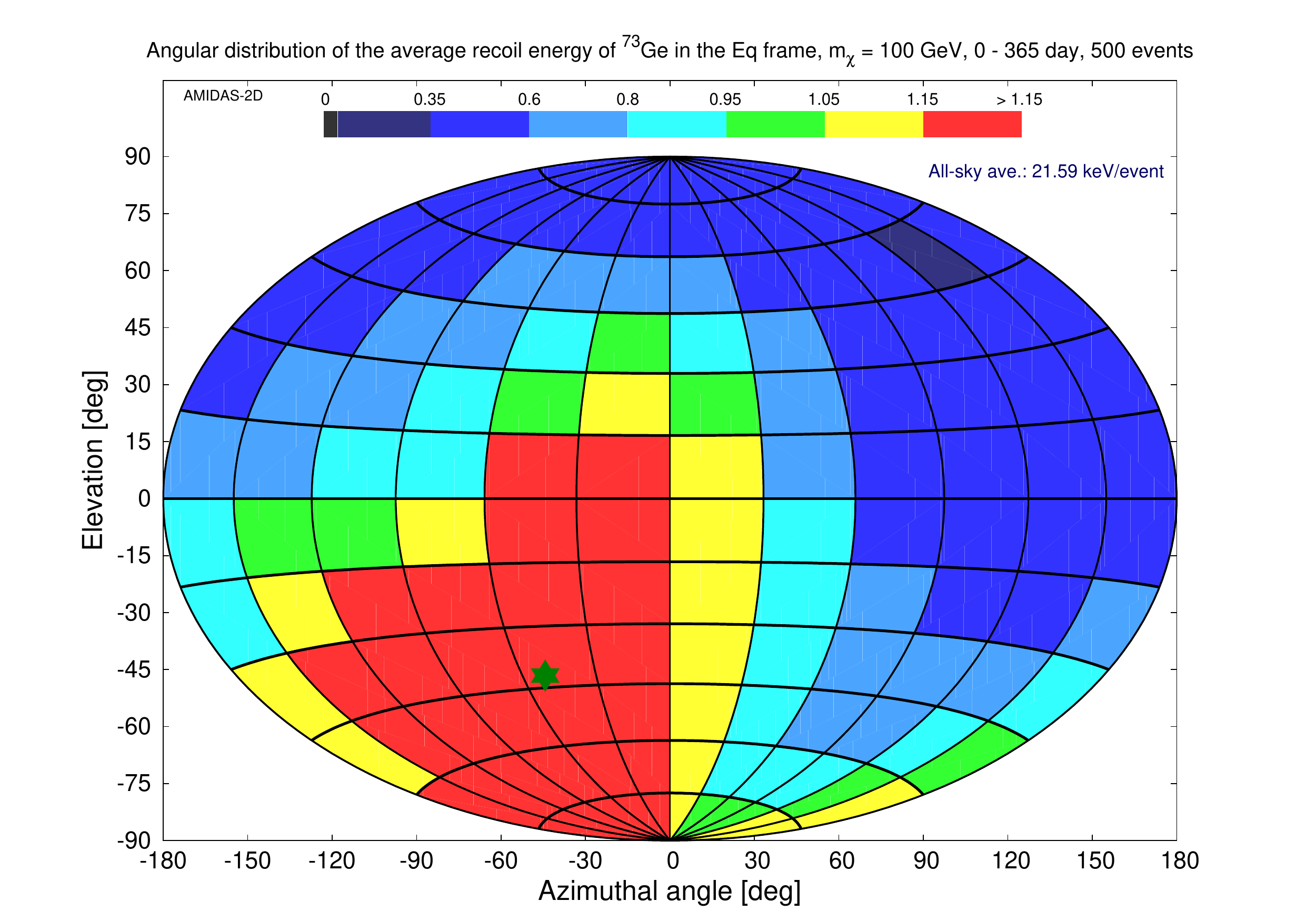}}%
 \caption{\footnotesize 100-GeV WIMPs off $\rmGe$}
 \end{subfigure}
 \begin{subfigure} [c] {5.5 cm}%
  \OnlinePlotNRangEq
   {QoN}
   {Xe129}
   {\includegraphics [width = 5.5 cm]
     {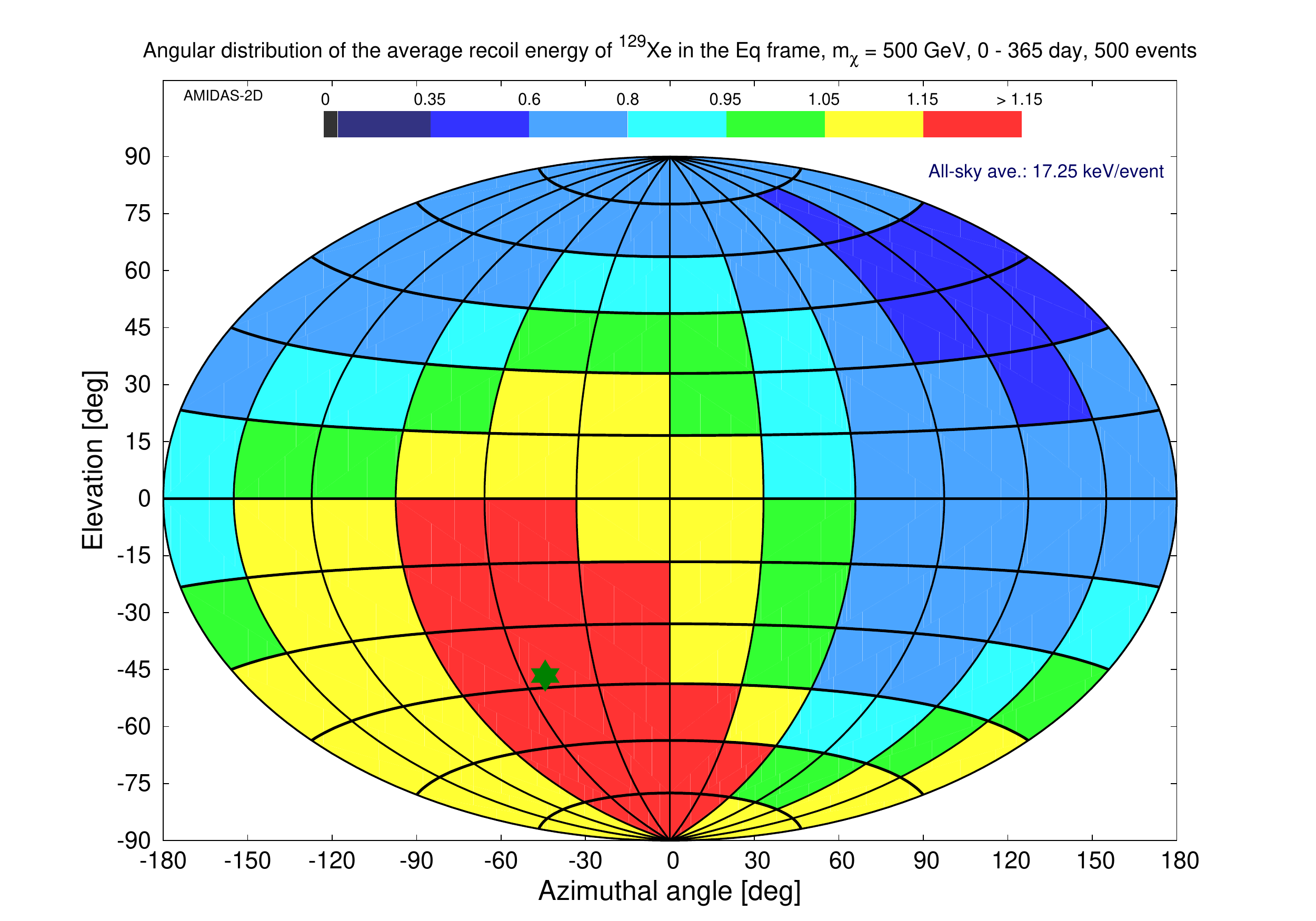}}%
 \caption{\footnotesize 500-GeV WIMPs off $\rmXe$}
 \end{subfigure}
\end{center}
\caption{
 Corresponding to Figs.~\ref{fig:NR_ang-Eq-0500-00000}:
 the angular distributions of
 the average recoil energy (per event)
 observed in the Equatorial coordinate system
 (in unit of the all--sky average values).
 Note that
 the all--sky average values
 depend on the WIMP mass and the target nucleus
 and thus
 are different in three plots.
 (Figures from Refs.~\cite{DMDDD-NR, AMIDAS-2D-web}).
}
\label{fig:QoN_ang-Eq-0500-00000}
\end{figure}

 In Figs.~\ref{fig:NR_ang-Eq-0500-00000}
 and \ref{fig:QoN_ang-Eq-0500-00000},
 we show
 the angular distributions of
 the WIMP--induced nuclear recoil flux
 and the corresponding average recoil energy (per event)
 observed in the Equatorial coordinate system
 (in unit of the all--sky average values),
 respectively.
 WIMPs of three different masses
 scatter off three different target nuclei
 have been presented:
 (a)  20-GeV WIMPs off $\rmF$,
 (b) 100-GeV WIMPs off $\rmGe$,
 (c) 500-GeV WIMPs off $\rmXe$.
 500 accepted
 scattering events on average
 (Poisson distributed)
 in one entire year
 have been recorded%
\footnote{
 Interested readers can click each plot
 in Figs.~\ref{fig:NR_ang-Eq-0500-00000}
 and \ref{fig:QoN_ang-Eq-0500-00000}
 to open the corresponding webpage of
 the animated demonstration
 with varying WIMP masses
 (for more considered target nuclei).
}.

 It can be seen clearly that
 the angular recoil--flux distributions of
 all three considered WIMP mass--target nucleus combinations
 would not (exactly) center
 at the theoretical direction of
 the WIMP wind
 (the dark--green star).
 The heavier the incident WIMPs and/or the target nucleus,
 the larger the northern deviation of
 the peak of
 the nuclear recoil flux.
 In constrast,
 the angular distributions of
 the average recoil energy of
 all three considered WIMP mass--target nucleus combinations
 seem to peak approximately
 around the theoretical WIMP--wind direction,
 with nevertheless
 some small (but observable)
 WIMP--mass and target dependent pattern differences.

 More detailed discussions about
 the angular distributions of
 the recoil flux and the (average) recoil energy of
 the WIMP--scattered target nuclei
 in different celestial coordinate systems
 can be found
 in Refs.~\cite{DMDDD-NR, DMDDD-NR-TAUP2021}.
 And
 in Ref.~\cite{DMDDD-v_theta},
 we will present
 the angular recoil--flux distributions
 with different WIMP mass--target nucleus combinations
 in different small (a few keV) recoil energy windows.

\subsection{3-D WIMP (effective) velocity distributions}
\label{sec:fv_eff}

 At the end of this section,
 we would like to mention that
 a technical difficulty
 in using the double differential event rate (\ref{eqn:d2RdQdOmega_SISD})
 would be about an analytic functional form of
 the 3-D WIMP velocity distribution
 in more realistic (and thus complicated) models of
 the Dark Matter halo,
 in particular,
 when one wants to include Dark Matter stream(s)
 \cite{OHare14}.

 Moreover,
 as demonstrated in Ref.~\cite{DMDDD-fv_eff}
 and mentioned in Sec.~\ref{sec:f1v_eff},
 due to the combination of
 the flux proportionality to the WIMP incident velocity
 and the (target and WIMP--mass dependent)
 nuclear form factor suppression,
 one needs to consider
 the 3-D ``effective'' velocity distribution of halo WIMPs
 scattering off target nuclei,
 even when
 the simplest isotropic (Maxwellian) velocity distribution
 in the Galactic coordinate system
 is adopted.
 This makes the situation
 much more complicated.

\section{Summary}

 In this paper,
 we discussed
 some unusual thoughts on
 (the incompleteness of)
 the expressions for
 the (double) differential event rates
 for elastic WIMP--nucleus scattering
 used in (directional) direct Dark Matter detection physics
 by comparing
 the commonly used expressions
 with those used
 in our double Monte Carlo scattering--by--scattering simulation procedure.

 Considering an incident WIMP
 moving with a given incident velocity
 and scattering off a target nucleus,
 one can find that
 the smaller the recoil angle,
 since the larger the corresponding recoil energy
 and thus the stronger the nuclear form factor suppression,
 the smaller the scattering probability could be.
 More precisely,
 the recoil--angle dependence of
 the differential scattering cross section
 with respect to the differential recoil angle
 indicates that
 head--on (zero--recoil--angle) WIMP--nucleus scattering
 should be impossible.
 Instead,
 WIMP signals with large recoil angles
 should be observed more frequently
 than we though earlier
 and,
 caused by the nuclear form factor suppression,
 the ratio of large--recoil--angle events
 should increase with the increasing target and/or WIMP mass.

 On the other hand,
 regarding the use of
 the double differential event rate
 in directional detection physics,
 we argued and demonstrated at first that,
 although
 the incident WIMP flux centers approximately
 at the opposite direction of the Solar Galactic movement,
 its decreasing distribution
 would not be ring--like around the center
 but rather distorted.
 Moreover,
 combined with
 the large--angle deflection of
 the WIMP--induced nuclear recoils,
 our numerical simulations
 show clearly that
 the angular distribution of
 the recoil flux of
 the WIMP--scattered target nuclei
 should be more complicated than
 the conventionally--believed
 ring--like--decreased distribution
 centered around the direction of
 the theoretically predicted WIMP wind.
 Once our target nuclei and the mass of incident WIMPs
 are pretty heavy,
 the angular recoil--flux distribution
 could be flattened (pretty) widely
 and the maximum could shift away (largely)
 from the WIMP--wind direction.
 Detailed investigations on
 the (target and WIMP--mass dependent)
 incident velocity--recoil angle distribution for
 elastic WIMP--nucleus scattering
 as well as
 the corresponding angular recoil--flux distributions
 will be announced soon.

 In summary,
 several not--frequently mentioned (but important) issues
 in (directional) direct DM detection physics
 have been argued and demonstrated in this work in detail.
 Unfortunately,
 we have so far no concrete suggestions
 for their alterations.
 We hope nevertheless that
 this work
 could initiate our colleagues
 to reconsider
 our theories/assumptions
 as well as
 earlier works in this field
 and eventually
 find out suitable improvements.

\subsubsection*{Acknowledgments}

 The author would like to thank
 the pleasant atmosphere of
 the Cancer Center of
 the Kaohsiung Veterans General Hospital,
 where part of this work was completed.
 This work
 was strongly encouraged by
 the ``{\it Researchers working on
 e.g.~exploring the Universe or landing on the Moon
 should not stay here but go abroad.}'' speech.

%
%
%
 %
%

%
%

%

%
%
%
\end{document}